\documentclass[review]{elsarticle}
\makeatletter
\def\ps@pprintTitle{
	\let\@oddhead\@empty
	\let\@evenhead\@empty
	\def\@oddfoot{\reset@font\hfil\thepage\hfil}
	\let\@evenfoot\@oddfoot
}
\makeatother

 \biboptions{comma,sort&compress}

\usepackage[table]{xcolor}

\usepackage{graphicx}
\usepackage{amsmath}
\usepackage{here}
\usepackage{cuted}

\usepackage{here}
\usepackage[dvips]{epsfig}
\definecolor{bluette}{rgb}{.2,.4,0}
\definecolor{salmon}{rgb}{.9,0.68,0.5}
\definecolor{motive}{rgb}{0.2,1,.5}
\definecolor{list}{rgb}{0.3,.8,.1}
\definecolor{moe}{rgb}{1,.7,.5}
\definecolor{mote}{rgb}{.7,.5,.6}
\definecolor{pisello}{rgb}{.1,1,0}
\definecolor{orange}{rgb}{1,.7,0}
\definecolor{oliva}{rgb}{.1,.5,0.3}
\definecolor{greenda}{rgb}{0,.3,.2}
\definecolor{greenli}{rgb}{0.5,.8,.0}
\definecolor{blueda}{rgb}{0,.1,.6}
\definecolor{purple}{rgb}{.7,.1,.2}
\definecolor{marrone}{rgb}{1,0.7,0}
\definecolor{pinky}{rgb}{1,0.8,0.8}
\definecolor{rose}{rgb}{1,0.4,0}

\def\beq{\begin{equation}}
\def\eeq{\end{equation}}
\def\bea{\begin{eqnarray}}
\def\eea{\end{eqnarray}}
\def\bq{\begin{quote}}
\def\eq{\end{quote}}

\def\nnb{\nonumber}
\def\ga{\left(}
\def\dr{\right)}

\def\rar{\rightarrow}
\def\lrar{\Longrightarrow}

\def\nnb{\nonumber}

\def\la{\langle}
\def\ra{\rangle}

\def\ba{\vspace*{-0.2cm}\begin{array}}
\def\ea{\end{array}\vspace*{-0.2cm}}

\def\b{$\bullet~$}
\def\d{$\diamond~$}

\def\als{\alpha_s}

\def\gg2{\la\alpha_s G^2 \ra}
\def\gg3{g^3f_{abc}\la G^aG^bG^c \ra}
\def\ggg4{\la\als^2G^4\ra}

\def\gg{\lag g^{2}_{s} G^2 \rag}
\def\ggg{\lag g^{3}_{s}G^3\rag}

\usepackage{amsmath}
\usepackage{slashed}
\usepackage{color}

\begin{document}

\markboth{Stephan Narison, Montpellier (FR)}{ }
\begin{frontmatter}

\title{ 
Revisiting the Muon Anomaly 
from $e^+ e^-\to$ Hadrons\,\tnoteref{invit}}
\tnotetext[invit]{Talk given at QCD25 - 40th anniversary of the QCD-Montpellier Conference. }
\author{Stephan Narison
}
\address{Laboratoire
Univers et Particules de Montpellier (LUPM), CNRS-IN2P3 and Univ. Montpellier, \\
Case 070, Place Eug\`ene
Bataillon, 34095 - Montpellier, France\\
and\\
Institute of High-Energy Physics of Madagascar (iHEPMAD)\\
University of Antananarivo, Ankatso 101, Madagascar}
\ead{snarison@yahoo.fr}


\date{\today}
\begin{abstract}

In this talk, I revisit and present a more comprehensive and systematic estimate  of the lowest order Hadronic Vacuum Polarization (HVP) contribution $a_\mu\vert_{hvp}^{lo}$ to the muon anomalous magnetic moment (muon anomaly) from $e^+e^-\to$ Hadrons obtained recently  in Ref.\,\cite{SNe}. 
New CMD-3 data are used for $e^+e^-\to 2\pi$ \,\cite{CMD3}  while
precise BABAR\,\cite{BABAR} and recent BELLE2\,\cite{BELLE} $e^+e^-\to 3\pi$ data are used
 to update the estimate of the $I=0$ isoscalar channel below the $\phi$-meson mass. 
 Adding the data compiled by PDG22\,\cite{PDG} above 1 GeV and the
QCD improved continuum used in Ref.\,\cite{SNe}, one deduces\,: $a_\mu\vert^{hvp}_{lo}=(7094\pm 37)\times 10^{-11} $.  A comparison with previous data driven ($e^+e^-$ and $\tau$-decays) estimates is done.
 Adding the  Higher Order $a_\mu\vert_{hvp}^{ho}$ corrections, the phenomenological estimate of the Hadronic Light by Light scattering up to NLO and  the QED and Electroweak (EW) contributions, one obtains\,: $\Delta a_\mu^{pheno}\equiv a_\mu^{exp}-a_{\mu}^{pheno}=  (81\pm  41)\times 10^{-11}$ where the recent experimental value $a_\mu^{exp}$ from Ref.\,\cite{FNAL1}  has been used.  This result  consolidates the previous one in Ref.[1], after adding the $\pi^0\gamma,\eta\gamma$ contributions and can be compared with the one from the most precise Lattice result $\Delta a_\mu^{lattice}= (90\pm 56)\times 10^{-11}$. Then, we deduce the (tentative) SM prediction average : $\Delta a_\mu^{SM}= (87\pm 33)\times 10^{-11}$. We complete the paper by revising our predictions on the LO HVP contributions in adding the $\pi^0\gamma,\eta\gamma$ contributions to the ones in Ref.\,\cite{SNe}. Then, we obtain: $a_\tau\vert^{hvp}_{lo}=(3516\pm 25)\times 10^{-11} $ and  $\Delta \alpha^{(5)}_{had}(M_Z^2)=(2770.7\pm 4.5)\times 10^{-5}$ for 5 flavours.
\end{abstract}
\end{frontmatter}
\newpage
\section{Introduction and Motivation}
\vspace*{-0.2cm}
Due to the (apparent) absence of direct productions of Particles beyond the SM at LHC, some Low-Energy precision Tests of the Standard Model (SM) predictions are hoped to detect some deviations from the SM predictions from virtual manifestations of new particles. 

Among these different processes, the measurement of the muon anomalous magnetic moment $a_\mu\equiv \frac{1}{2}(g-2)_\mu$  (called hereafter muon anomaly) seems to be promising for detecting some physics beyond the SM predictions.

The new precise experimental results from E821(BNL)\,\cite{BNL}  and FNAL\,\cite{FNAL1}   have motivated  improved theoretical estimate  (data driven, lattice calculations) of the Lowest Order (LO) - Hadronic Vacuum Polarization (HVP) which is the main source of the theoretical errors on $a_\mu$.  The aim of this paper is to improve and review the estimate of the LO - HVP using $e^+e^-\to$ Hadrons data in the whole time-like region and to present  the new status of $a_\mu$ from SM using $e^+e^-$ data driven.

\section{Historical measurements of the muon anomaly}
\section*{\hspace*{0.5cm} \b The pionniering experiments (1957-1965)}
\d Since the pioneering work of Ref.\,\cite{CASSEL} using a polarized 110 MeV postive muon beam produced from pion decays who found\,:
\beq
a_\mu^{exp}= (2\pm 7)\times 10^{-4},
\eeq

\d Ref.\,\cite{COFFIN} uses a magnetic resonance technique on the  CHBr3 target and obtained\,:
\beq
a_\mu^{exp}= (26\pm 9)\times 10^{-4},
\eeq
compared to the QED prediction of $12\times 10^{-4}$.

\d A new experiment  using a precession technique for measuring the ratio of the muon precession frequency to that of the proton in the same magnetic field has been used. In this way, Ref.\,\cite{GARWIN} found the lower limit:
\beq
a_\mu^{exp}\geq (122\pm 8)\times 10^{-5},   
\eeq
based on a  lower limit of the muon mass. 

\d  Later on, Ref.\,\cite{CHARPAK} uses a  precession for 100 MeV/c muons as a function of storage time t and found:
\beq
a_\mu^{exp}=(1162\pm5)\times 10^{-6},
\eeq
compared to the QED prediction to order $(\alpha/\pi)^2$ : 
\beq
a_\mu^{qed}=1165\times 10^{-6} 
\eeq
\section*{\hspace*{0.5cm} \b The CERN muon storage ring (1961-1976)}
Since then, a muon storage ring dedicated to a direct measurement of $a_\mu$ via the produced muon precessing spin has been built at CERN. The final report on anomalous magnetic moments of positive and negative muons are found to be\,\cite{BAILEY}\,:
\beq
a_\mu^+= (1165911\pm 11) \times 10^{-9}~~~~~{\rm and} ~~~~~ a_\mu^- = (1165937\pm 12) \times 10^{-9}
\eeq
leading to the average:
\beq
\la a_\mu\ra = (1165924.0\pm 8.5) \times10^{-9},
\eeq
in good agreement with the theoretical estimate within the Standard Model \,\cite{CALMET}\,:
\beq
a_\mu^{th} = (1165920.6 \pm 12.9) \times10^{-9}.
\eeq
\section*{\hspace*{0.5cm} \b Improved measurements of $a_\mu$ from E821 (BNL) and FNAL}
Later on, new dedicated experiments for improving the measurement of the muon anomaly have been built. 

\d  The E821 at Brookhaven (1997-2001) uses the same techniques as the CERN muon storage ring but with innovative technologies (continuous superconducting magnet, having high
field uniformity, continuous current rather than pulsed,...). The  final report of the E821 (BNL) series of precise measurement  leads to the value\,\cite{BNL}:
\beq
a_\mu\vert^{exp}_{bnl} = (116592080\pm 63)\times 10^{-11}, 
\label{eq:bnl}
\eeq
which is about $(2.2\sim 2.7)\sigma$ above the SM predictions compiled in\,\cite{BNL}.

\d The recent measurement of $a_\mu$ from FNAL Runs 1 to 6 \,\cite{FNAL1} \,:
\beq
a_\mu\vert^{exp}_{fnal} = (116592070.5\pm 14.8)\times 10^{-11},
\label{eq:final}
\eeq
  \vspace*{-0.5cm}
\begin{figure}[hbt]
\begin{center}
\includegraphics[width=6cm]{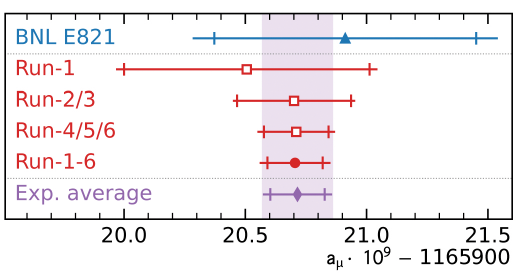}
\vspace*{-0.5cm}
\caption{\scriptsize $a_\mu$ from BNL and FNAL }\label{fig:fnal}
\end{center}
\vspace*{-0.5cm}
\end{figure}  

has improved the precision of the previous E821 (BNL) value in Eq.\,\ref{eq:bnl} by a factor about\,4
and  the previous FNAL Run2 value\,\cite{FNAL2}:
\beq
a_\mu\vert^{exp}_{fnal} = (116592057\pm 25)\times 10^{-11},
\eeq
by a factor of 1.7.  The previous results lead to the BNL and FNAL average\,(see Fig\,\ref{fig:fnal} from \cite{FNAL1})\,:
\beq
\la a_\mu^{exp}\ra = (116592071.5\pm 14.5)\times 10^{-11}.
\eeq

\section{The QED and EW contributions to  $a_\mu$}
The previous experimental progresses stimulate an improvement of the theoretical predictions.

\section*{\hspace*{0.5cm} \b The QED contributions to  $a_\mu$}
\begin{figure}[hbt]
\begin{center}
\includegraphics[width=8cm]{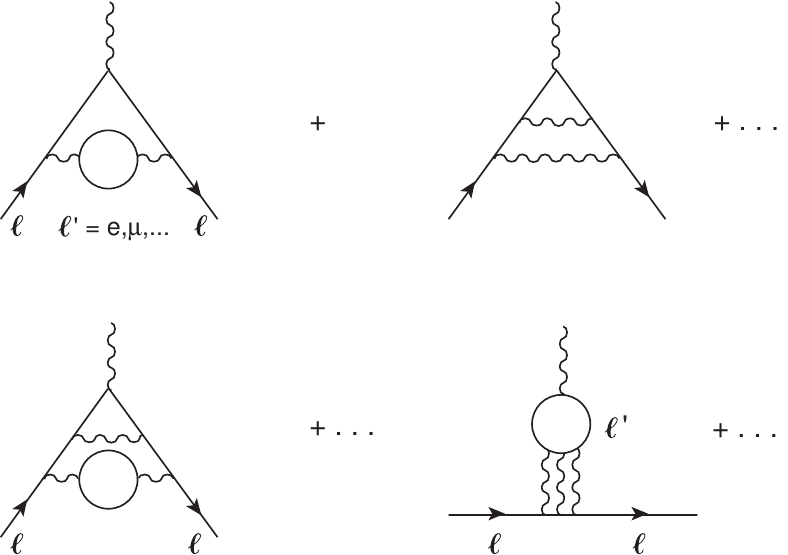}
\caption{\scriptsize NLO and N2LO  QED quantum corrections to $a_\mu$ }\label{fig:amu-ho}
\end{center}
\vspace*{-0.5cm}
\end{figure}

Since the lowest order Schwinger quantum correction from diagram with an exchanged  virtual photon\ (see Fig.\ref{fig:amu-lo})\,:
\begin{figure}[hbt]
\begin{center}
\includegraphics[width=4cm]{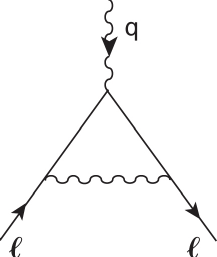}
\caption{\scriptsize Lowest order quantum correction to $a_\mu$ }\label{fig:amu-lo}
\end{center}
\vspace*{-0.5cm}
\end{figure}  

\beq
a_\mu\vert^{qed}_{lo} =\frac{\alpha}{2\pi} 
\eeq
a large amount of efforts have been done for calculating Higher Order (HO) corrections (for reviews, see e.g.\,\cite{CALMET,DERAFAEL,AOYAMA,WP25}).
The Next-to-Leading (NLO) and Next-to-Next-to-Leading (N2LO) diagrams are shown for illustration in Fig.\,\ref{fig:amu-ho}. 
The QED contributions is known to 10th order $\alpha^5$ where at this order, the $a_\mu$ value is sensitive to the input value of $\alpha$, the uncertainties of the $\tau$-lepton,
of the 8th and 10th order contributions and on the estimate of the 12th order contribution. The uncertainty on $\alpha$ leads to a value of $a_\mu$ which differs by about 0.137$\times 10^{-11}$ and is about the estimate of the 12th order term. Therefore, Ref.\,\cite{WP25} takes the conservative value:
\beq
a_\mu^{qed}= (116 584 718.8\pm 2.0) \times 10^{-11}.
\eeq
\section*{\hspace*{0.5cm} \b The Electroweak (EW)  corrections to  $a_\mu$}
\begin{figure}[hbt]
\begin{center}
\includegraphics[width=9cm]{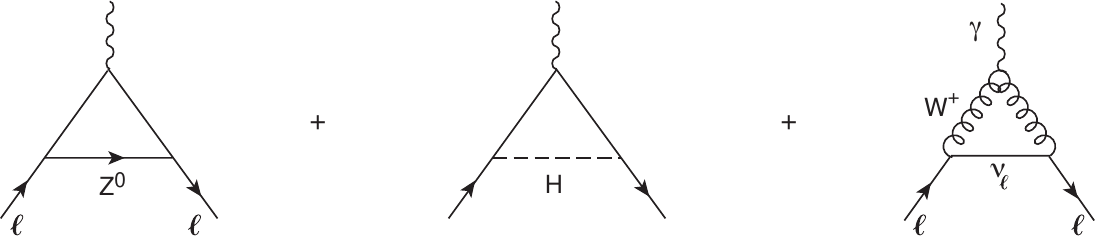}
\caption{\scriptsize LO electroweak corrections to $a_\mu$ }\label{fig:amu-ew}
\end{center}
\vspace*{-0.5cm}
\end{figure}

These corrections are illustrated to Lowest Order (LO)  by the diagrams in Fig.\,\ref{fig:amu-ew}
Up to two virtual fermion and boson loop corrections and including the leading-log three-loop corrections, one obtains\,\cite{WP25,AOYAMA}:
\beq
a_\mu^{ew}=  (154.4\pm 4.0) \times 10^{-11}.
\eeq
\section*{\hspace*{0.5cm} \b Conclusions from the QED and Electroweak (EW)  corrections to  $a_\mu$}
From the previous analysis, we deduce:
\beq
a_\mu^{qed}+a_\mu^{ew}= (16 584 873.2\pm 4.5) \times 10^{-11},
\eeq
which is 3.3 more precise than the recent experimental value in Eq.\,\ref{eq:final} but differs by:
\beq
\la a_\mu^{exp}\ra- a_\mu^{qed}-a_\mu^{ew} = (7197.3\pm 14.5_{exp}\pm 4.5_{th})\times 10^{-11}.
\label{eq:amuth}
\eeq
This discrepancy beween Experiment with QED $\oplus$ EW is expected to be compensated by the Hadronic contributions to $a_\mu$ and by some eventual contributions beyond the Standard Model (BSM). 

\section{The lowest order hadronic vacuum polarization contribution $a_\mu\vert^{hvp}_{lo}$}
\begin{figure}[hbt]
\begin{center}
\includegraphics[width=4cm]{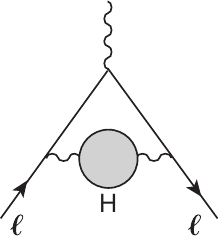}
\caption{\scriptsize Lowest order hadronic vacuum polarization contribution to $a_\mu$ }\label{fig:amu-hvp-lo}
\end{center}
\vspace*{-0.5cm}
\end{figure}

This contribution is given by the diagram in Fig.\,\ref{fig:amu-hvp-lo} where the QED lepton loop is replaced
by  hadrons and quark loops of QCD. 
This contribution can be written in terms of a convolution integral in terms of the spectral function of the electromagnetic 
two-point correlator which is connected to the total cross-section of the $e^+e^-\to$ Hadrons via the optical theorem
\,(for an historical review see Ref.\,\cite{DERAFAEL})\,:
\beq
a_\mu\vert_{lo}^{hvp}= \frac{1}{4\pi^3}\int_{4m_\pi^2}^\infty dt\, K_\mu(t)\sigma (e^+e^-\to {\rm hadrons})
\label{eq:amu-form}
\eeq
where $K_\mu(t)$ is the QED kernel function\,:
\beq
K_\mu(t)=\int_0^1dx\frac{x^2(1-x)}{x^2+(t/m_\mu^2)(1-x)}.
\label{eq:kmu}
\eeq
$K(t)$  behaves as $1/t$ at low-energy such that the integral is sensitive to the lowest meson mass contributions. 

In the following we shall discuss in details the different meson  contributions to $a_\mu\vert^{hvp}_{lo}$ through the $e^+e^-\to$ Hadrons 
total cross-section.  
\section*{\hspace*{0.5cm} \b The $\rho$ meson contribution below 0.99 GeV}
We are aware that there is a discrepancy among different data for the isovector channel $e^+e^-\to\pi^+\pi^-$.  Before a clarification of a such discrepancy, 
 we have choosen\,\cite{SNe} to work with  the recent CMD-3 data\,\cite{CMD3} in the region below 0.993 GeV. However, some recent different critical tests of the CMD-3 do not find any loophole from the analysis of the data\,\cite{WP25}. 
 We shall work with:
 \beq
 R^{ee} \equiv \frac{ \sigma (e^+e^-\to {\rm I=1\,hadrons})}{\sigma (e^+e^-\to \mu^+\mu^-)}, 
\eeq
which is related to the measured pion form factor $F_\pi$ as:
\beq
R^{ee}= 
\frac{1}{4}\ga 1-\frac{4m_\pi^2}{t}\dr^{3/2} \vert F_\pi\vert^2.
\eeq

\subsection*{\hspace*{1cm} \d Test of the Breit-Wigner parametrization of the pion form factor from $2m_\pi\to 0.88$ GeV}
\begin{figure}[hbt]
\begin{center}
\includegraphics[width=11.cm]{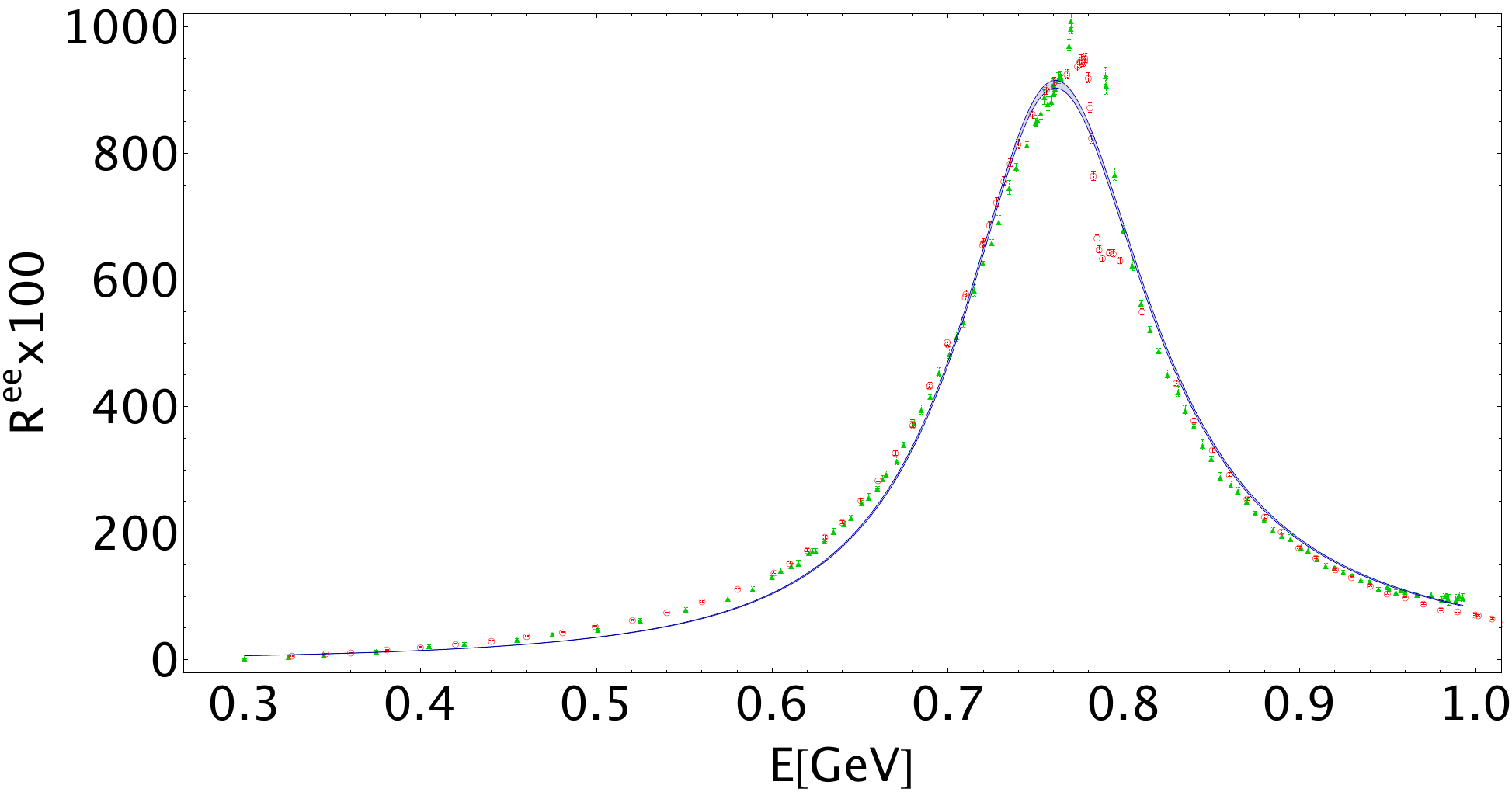}
\caption{\footnotesize  Fit of the data without $\omega$  using a minimal Breit-Wigner parametrization of the PDG\,\cite{PDG} compilation (green triangle) . A comparison with the new CMD-3\,\cite{CMD3} data is given (open red circle). . } \label{fig:rho}
\end{center}
\vspace*{-0.5cm}
\end{figure}

We isolate the $I=1$ part by subtracting  the $\omega$ meson contribution and the  shoulder around the peak 
due to the $\omega-\rho$ mixing. Then, we fit the data using the  optimized $\chi^2$ Mathematica program {\it Findfit}. In so doing we fit separately the maximum and minimum values of each data points. To minimize the number of the free parameters, we use as input the value $\Gamma(\rho\to e^+e^-)$ from PDG\,\cite{PDG}. Then, we obtain the parameters:
\bea
{\rm Set}\, 1:  && \Gamma(\rho\to e^+e^-)=(7.03+ 0.04)\,{\rm keV} \lrar  M_{\rho}=755.56\, {\rm MeV} ,~~~~~~~~\Gamma_\rho^{\rm tot}=  132.04\,\rm{MeV}.\nnb\\ 
{\rm Set}\, 2:  && \Gamma(\rho\to e^+e^-)=(7.03-0.04)\,{\rm keV}\lrar  M_{\rho}=755.65\, {\rm MeV} ,~~~~~~~~\Gamma_\rho^{\rm tot}=  131.92\,\rm{MeV},
\eea
where the mass slightly differs from  $M_\rho=775.5$ MeV quoted by PDG\,\cite{PDG} obtained using a more involved parametrization of the pion form factor and including $\rho-\omega$ mixing but it is satisfactory within our simple Breit-Wigner of the form factor. The result of the analysis is shown in Fig.\,\ref{fig:rho}.
\subsection*{\hspace*{0.5cm} \d  Fitting procedure}
 Then, we use the FindFit Mathematica program with optimised $\chi^2$. The data are parametrized using polynomials  or/and a simple Breit-Wigner. Our optimal result  is the mean value of these two extremal ones. We may expect that the results are more conservative but eventually less accurate that the one from a fit of the correlated data used currently in the literature.  Our fitting procedure will be extended to the different  available data until 11.5 GeV. 
\subsection*{\hspace*{0.5cm} \d  Improving the $\rho$ meson contribution below 0.993 GeV}
Though our minimal BW parametrization gives a quite good determination of the $\rho$-meson parameters compared to the PDG compilation, we can see in Fig.\,\ref{fig:rho} that there are regions which need to be improved. This is necessary for a high-precision determination of $a_\mu$. 
 In so doing , instead of being involved in the theoretical parametrization of the data (accuracy of chiral perturbaton theory, complicated Breit-Wigner parametrization of the pion form factor $F_\pi$,...),  we just fit the data by  dividing the region below 0.993 GeV into 5 subregions in units of GeV\,: 
\beq
[2m_\pi\to 0.6],\,\,\,[0.6\to 0.776],\,\,\,[0.776\to 0.786],\,\,\,[0.786 \to 0.810],\,\,\, [0.810\to .993],
\eeq
and apply the fitting procedure discussed in the previous subsection.  
The fit is done for the CMD-3 data (open red circle : purple region).   The PDG22 data (green triangle ; yellow region) are shown  for a comparison below the $\rho$ meson mass. To be conservative, we fit separately the upper and lower values of the data for a given experiment.  One can notice that in the regions from 0.4 to 0.6 GeV and 0.68 to 0.77 GeV, the CMD-3 data are systematically above the PDG22 compilation.  
 
\begin{figure}[hbt]
\begin{center}
\includegraphics[width=7.cm]{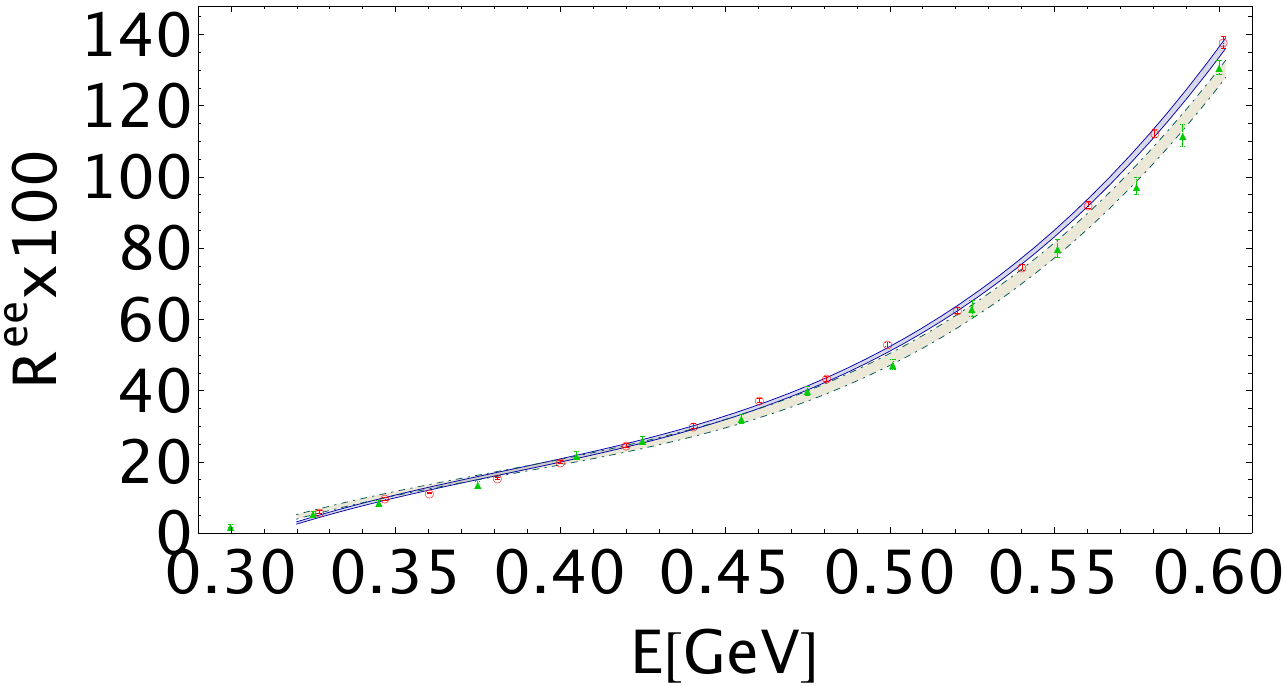}
\includegraphics[width=7.cm]{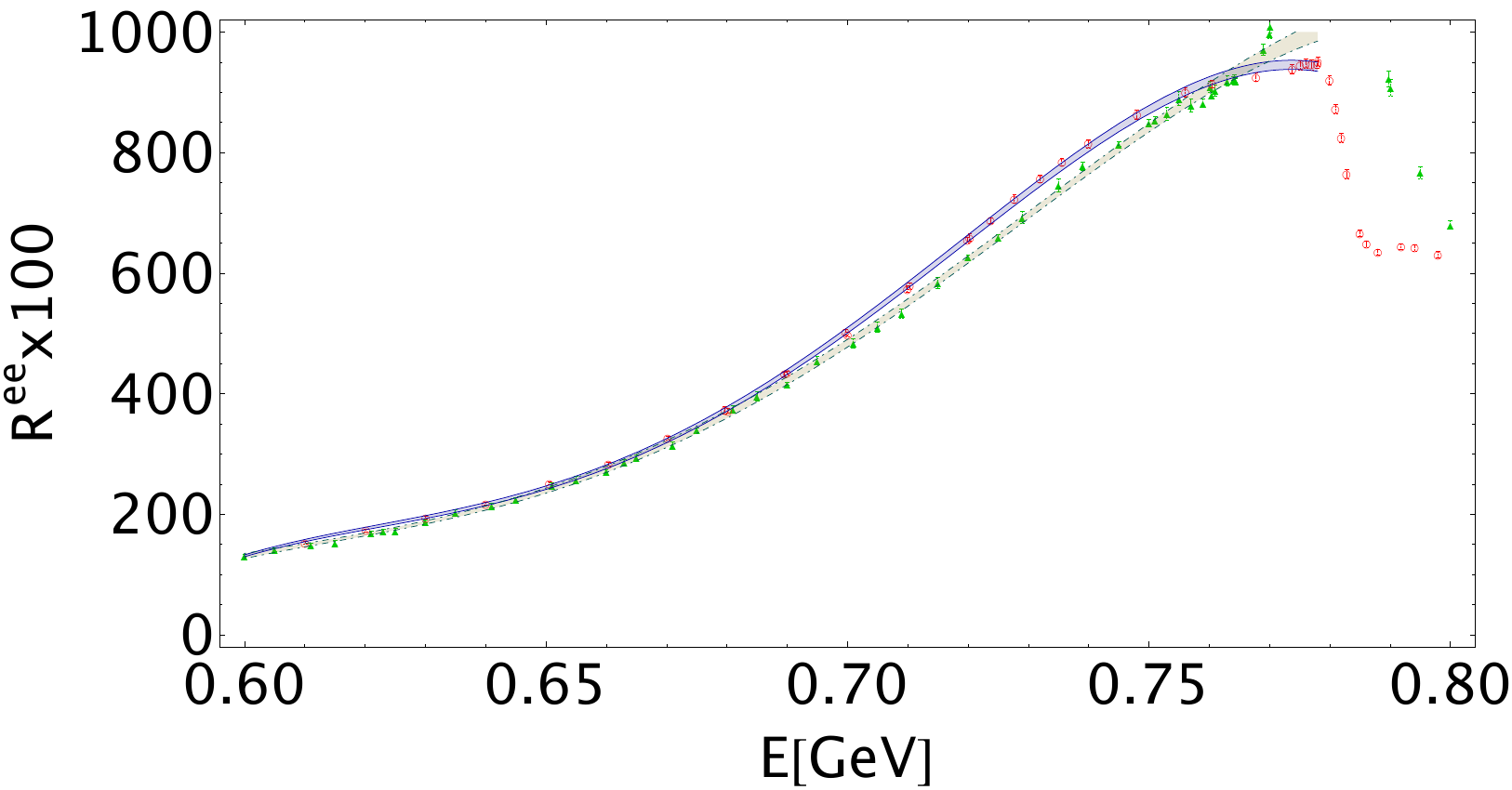}
\caption{\footnotesize  Fit of the data in the region $2m_\pi$ to 0.776 GeV. The purple curve is the fit of the CMD-3 data (open red circle). The PDG22 compilation is shown in green triangle (yellow region).} \label{fig:rho1}
\end{center}
\vspace*{-0.5cm}
\end{figure} 
\begin{figure}[H]
\begin{center}
\includegraphics[width=6cm]{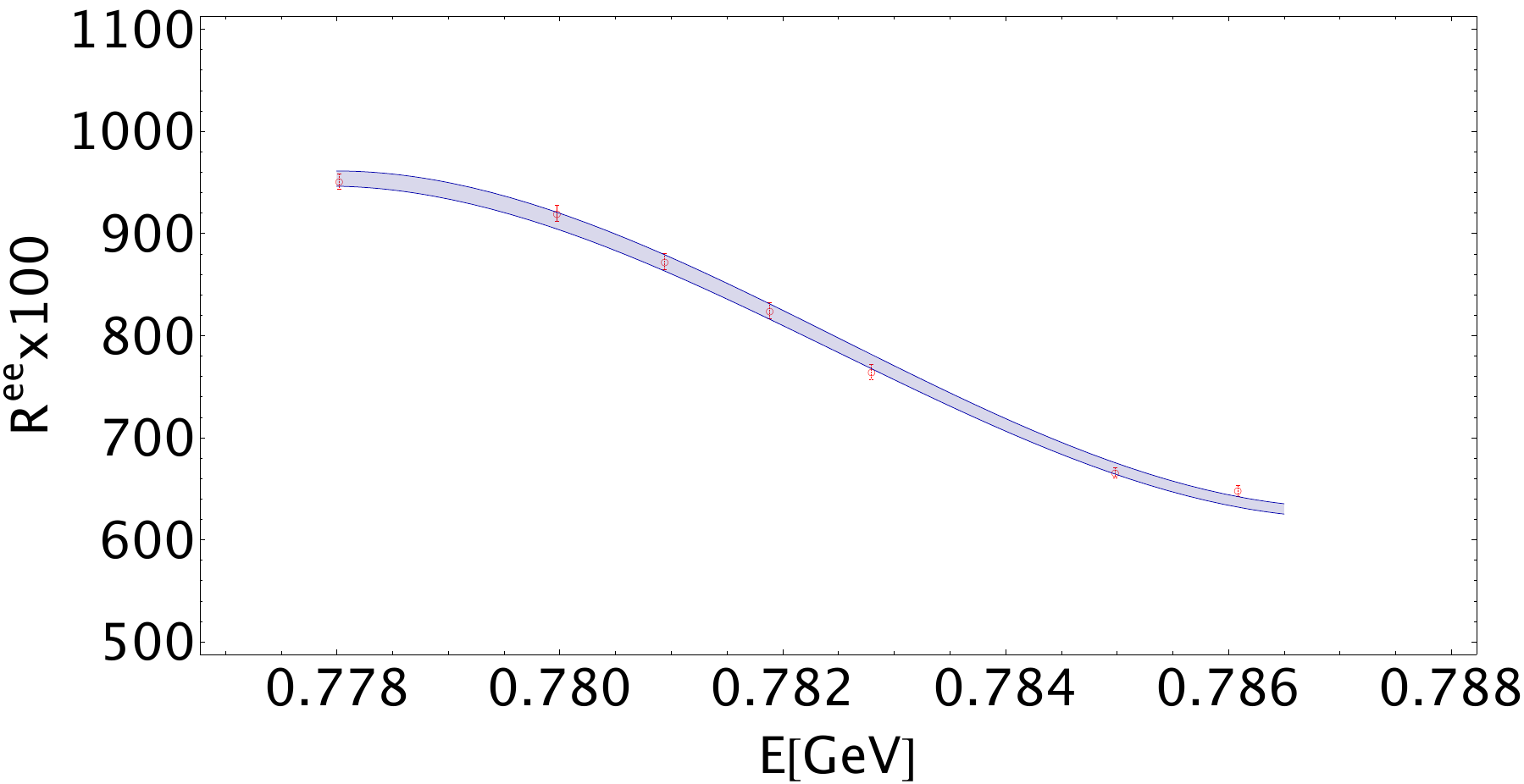}
\includegraphics[width=6cm]{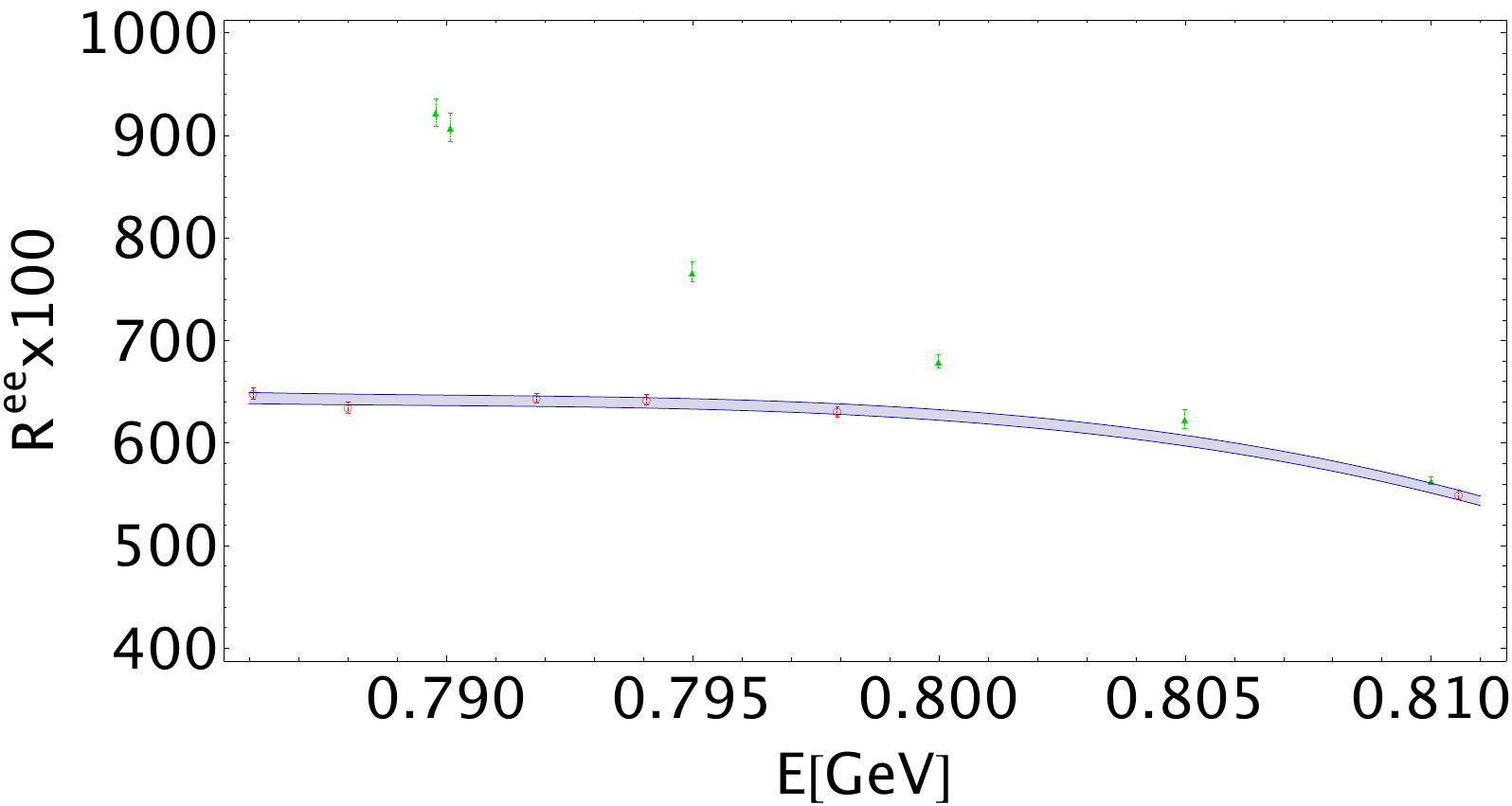}
\includegraphics[width=6cm]{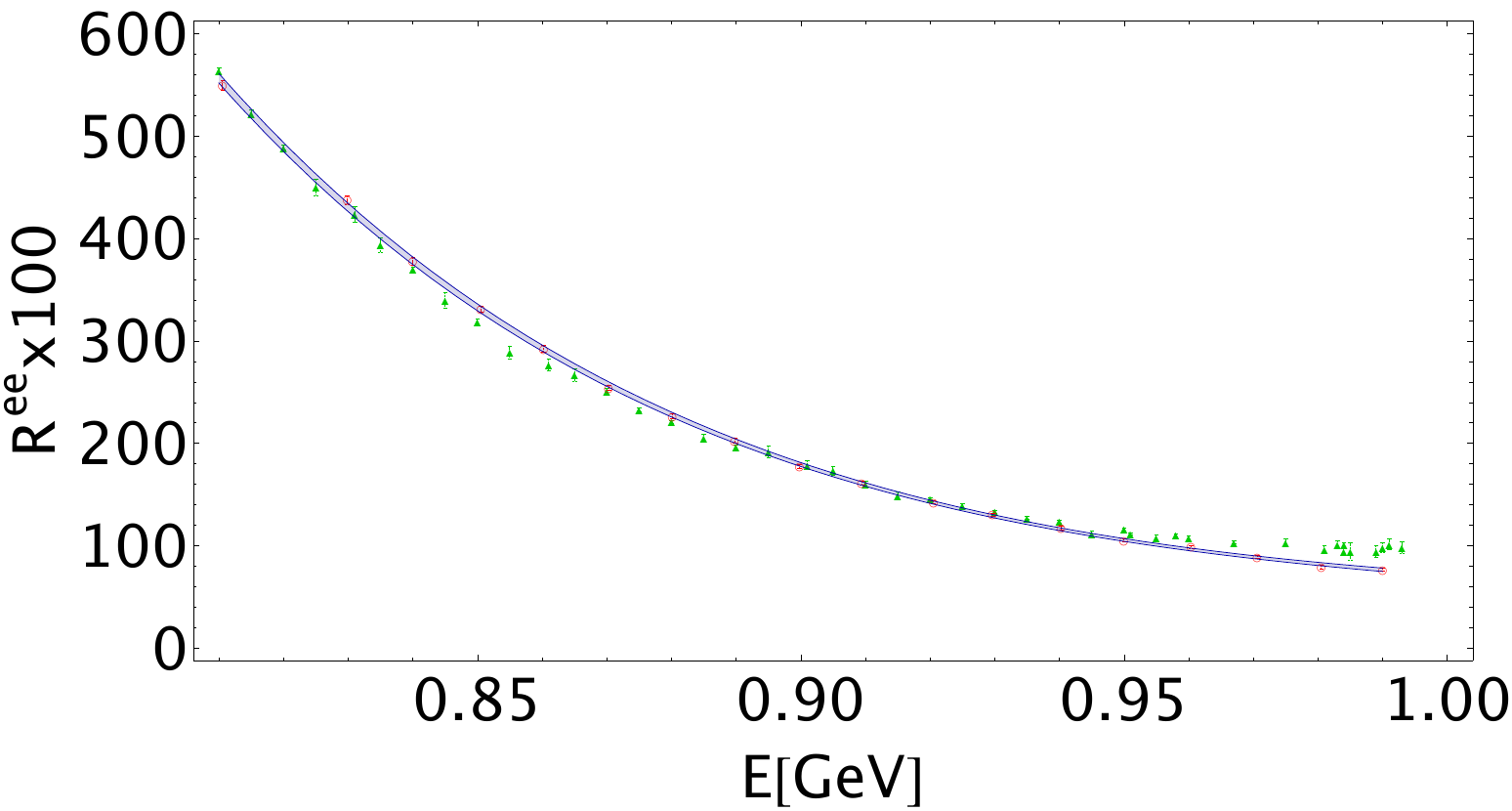}
\caption{\footnotesize  The same as in Fig.\,\ref{fig:rho1} but in the region 0.776 to 0.993 GeV. } \label{fig:rho2}
\end{center}
\vspace*{-0.5cm}
\end{figure} 
We cut the fit at 0.993 GeV where the ratio R starts to increase indicating the opening of new channels. 
Collecting the results from different regions, we deduce from the CMD-3 data\,:
\beq
a_\mu\vert_{lo}^{hvp}\vert_{cmd-3}(2m_\pi\rar 0.993) = (5109.3\pm 35.2)\times 10^{-11},
\eeq
where 76\% of the contribution comes from the region below the $\rho$-meson mass.
\section{Light $I=1$ isovector mesons  from 0.993 to  1.875 GeV}
\subsection*{\hspace*{0.5cm} \b  The region from 0.993 to 1.5 GeV}
\begin{figure}[hbt]
\begin{center}
\hspace*{0.5cm} {\bf a)} \hspace*{8cm} {\bf b)} \\
\includegraphics[width=8.cm]{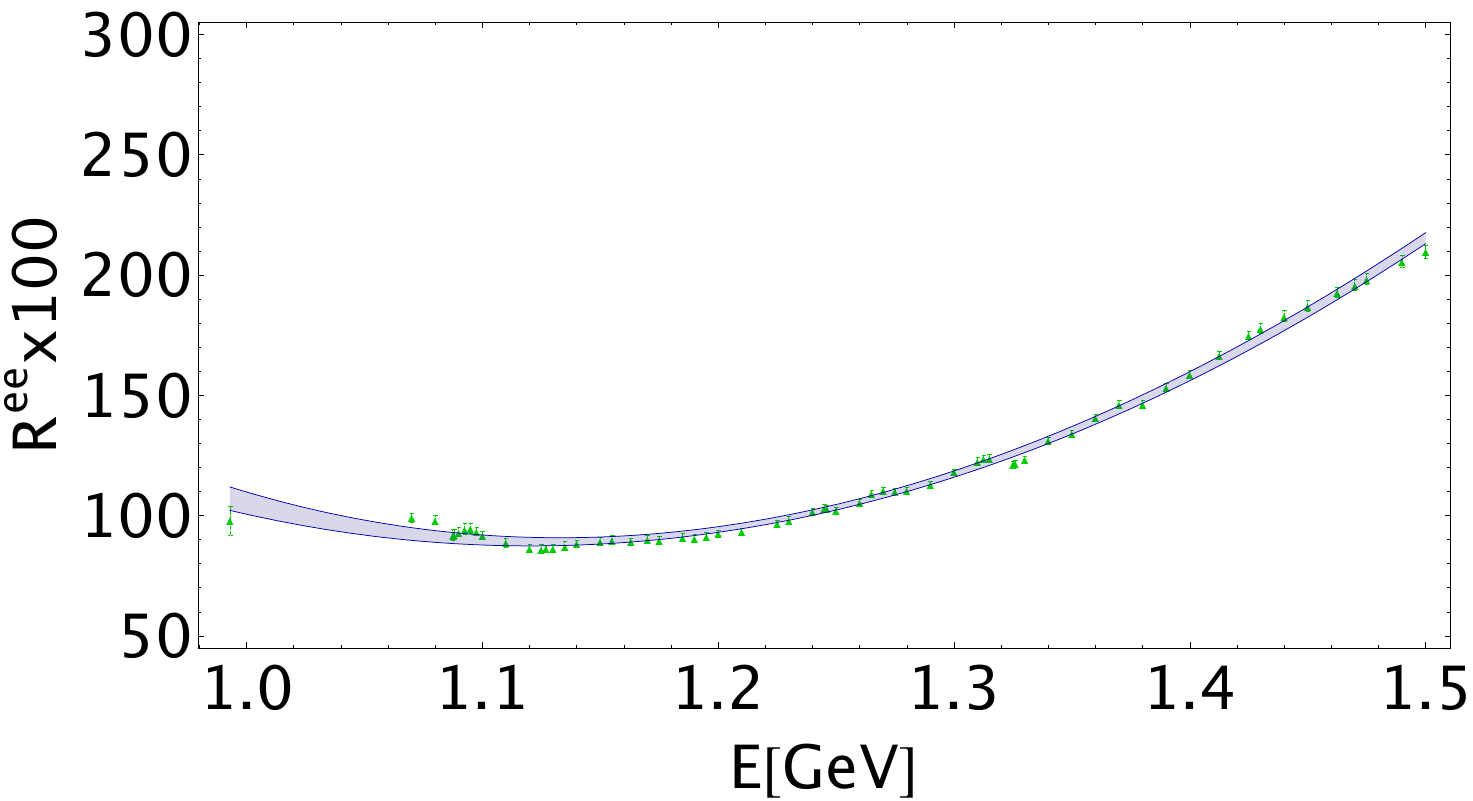}
\includegraphics[width=8.cm]{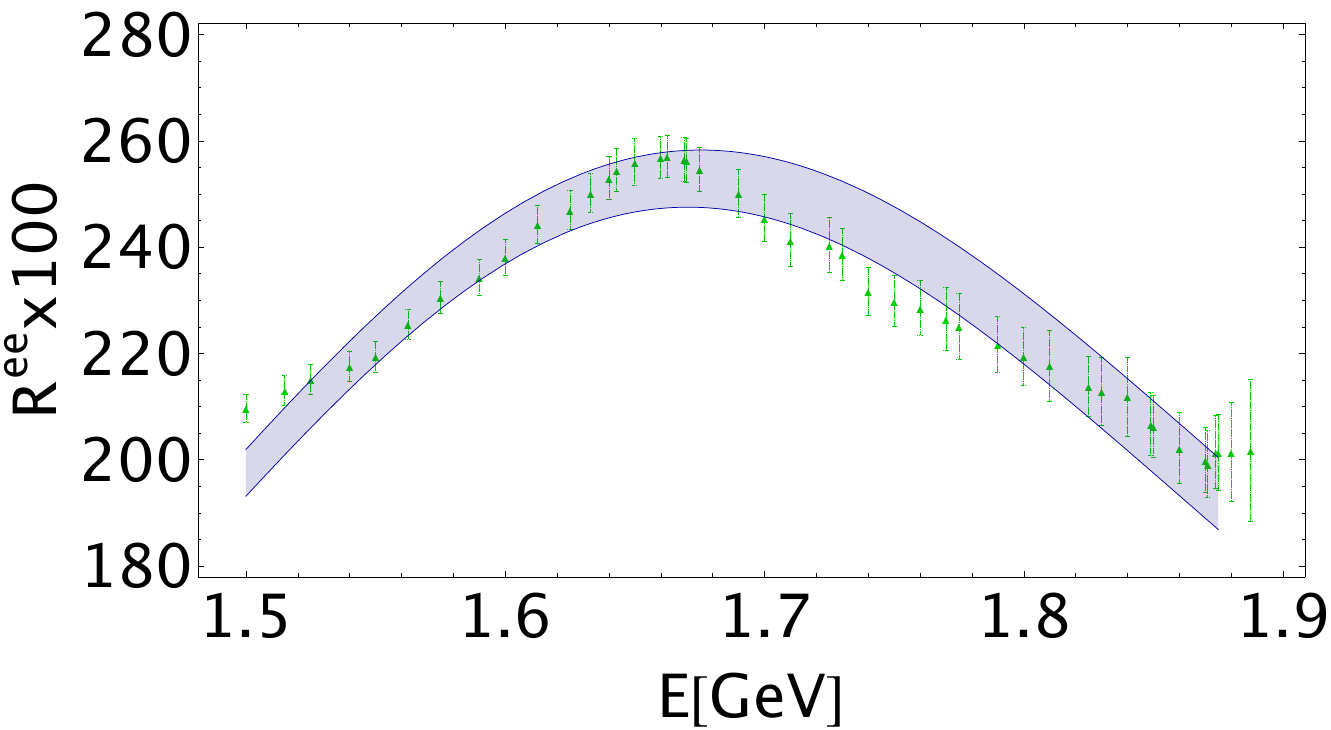}

\caption{\footnotesize  {\bf a)} Fit of the PDG data using a cubic polynomial  interpolating  formula; {\bf b)} Fit of the PDG data using a minimal Breit-Wigner parametrization.} \label{fig:3pi} \label{fig:rhoprime}
\end{center}
\vspace*{-0.5cm}
\end{figure}

We fit the data using a simple interpolation program with polynomials. We substract the $3\pi$ backgrounds by using the $SU(2)$ relation between the isoscalar and isovector states (a suppression 1/9 factor). 
We neglect the $\bar KK$ contributions from isoscalar sources which, in addition to the $SU(3)$ suppression factor is also suppressed by phase space. The fit is shown in Fig.\ref{fig:3pi}a).
We obtain:
\beq
a_\mu\vert_{lo}^{hvp}(0.993\rar 1.5) = (354.4\pm 6.7)\times 10^{-11},
\eeq

\subsection*{\hspace*{0.5cm} \b The $\rho'(1600)$ from 1.5 to   1.875 GeV}
 We use a Breit-Wigner fit for a $\rho'$ meson. We obtain :
\beq
M_{\rho'}=1.6\,{\rm GeV}\,\,\, {\rm and}\,\,\,\Gamma(\rho'\to e^+e^-) = [10.5\,({\rm resp.}\, 9.73)]~{\rm keV}~~~~~~~~\Gamma_{\rho'}^{\rm tot}=  [720\,({\rm resp.}\,694)]\,\rm{MeV},
\eeq
from the high (resp. low) data points. The fit is shown in Fig.\,\ref{fig:rhoprime}b).
We deduce:
\beq
a_\mu\vert_{lo}^{hvp}[\rho'(1600)] = (237.6\pm 5.7)\times 10^{-11},
\eeq

\section{Light $I=1$ isovector mesons  contributions from $2m_\pi$ to  1.875 GeV}
From the previous analysis, we deduce the total sum of the $I=1$ isovector meson contributions
from $2m_\pi$ to  1.875 GeV:
\beq
a_\mu\vert_{lo}^{hvp}(2m_\pi\rar 1.875) = (5701.3\pm 36.3)\times 10^{-11}.
\label{eqq:i=1}
\eeq
The previous different results are compiled in Table\,\ref{tab:amu1}. 
\section{The $\omega(780)$  $I=0$ isoscalar meson}
\subsection*{\hspace*{0.5cm} \d The $\omega(780)$ meson contribution below 0.96 GeV}

\begin{figure}[hbt]
\begin{center}
\includegraphics[width=7cm]{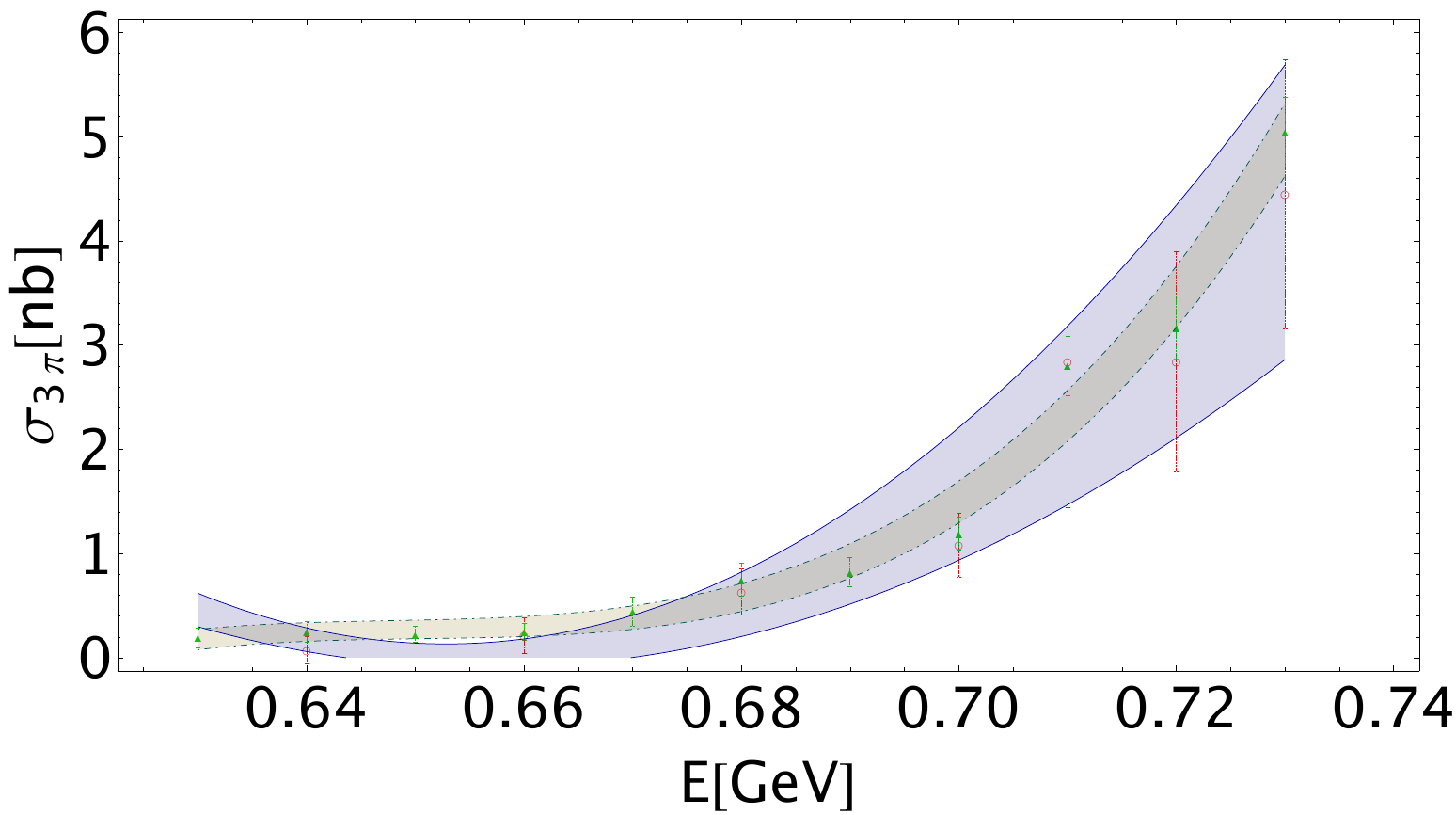}
\includegraphics[width=7cm]{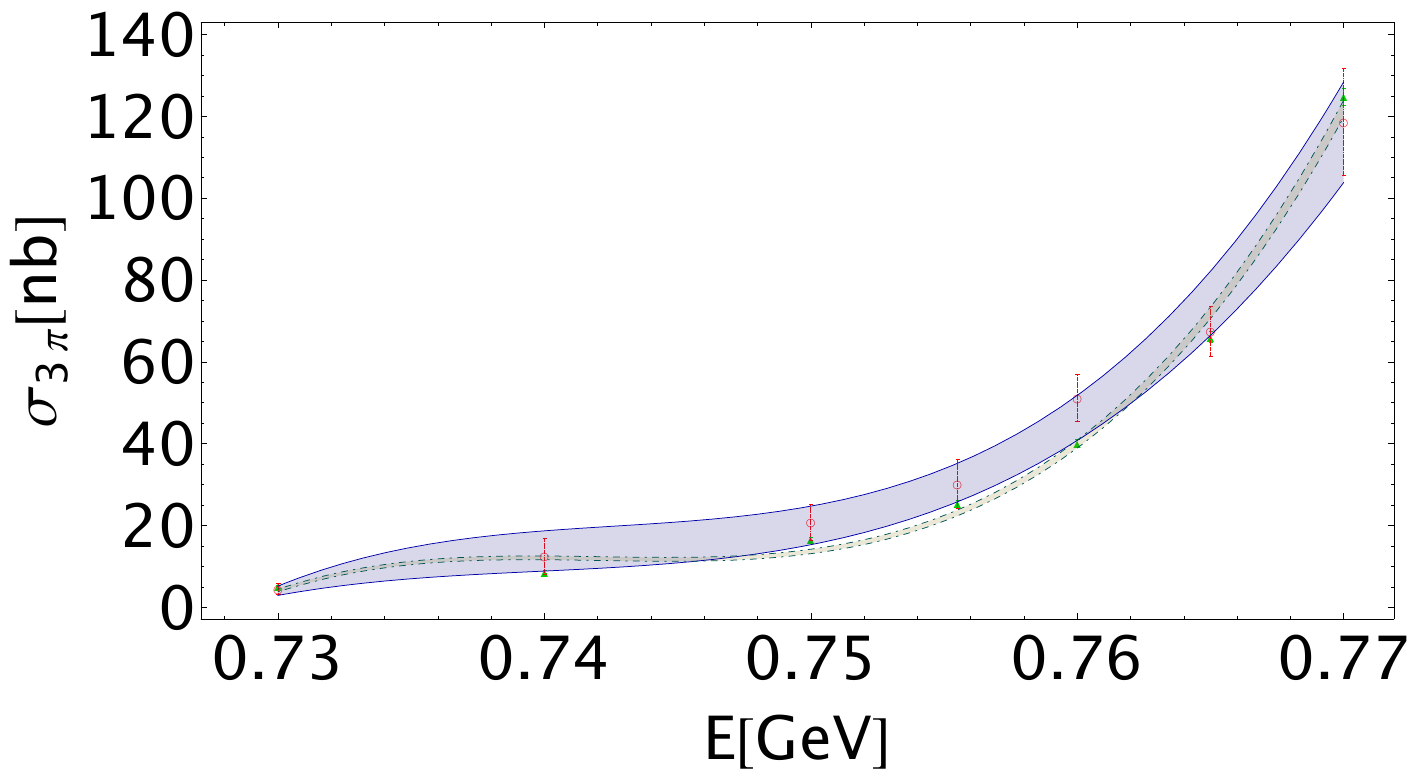}
\includegraphics[width=7cm]{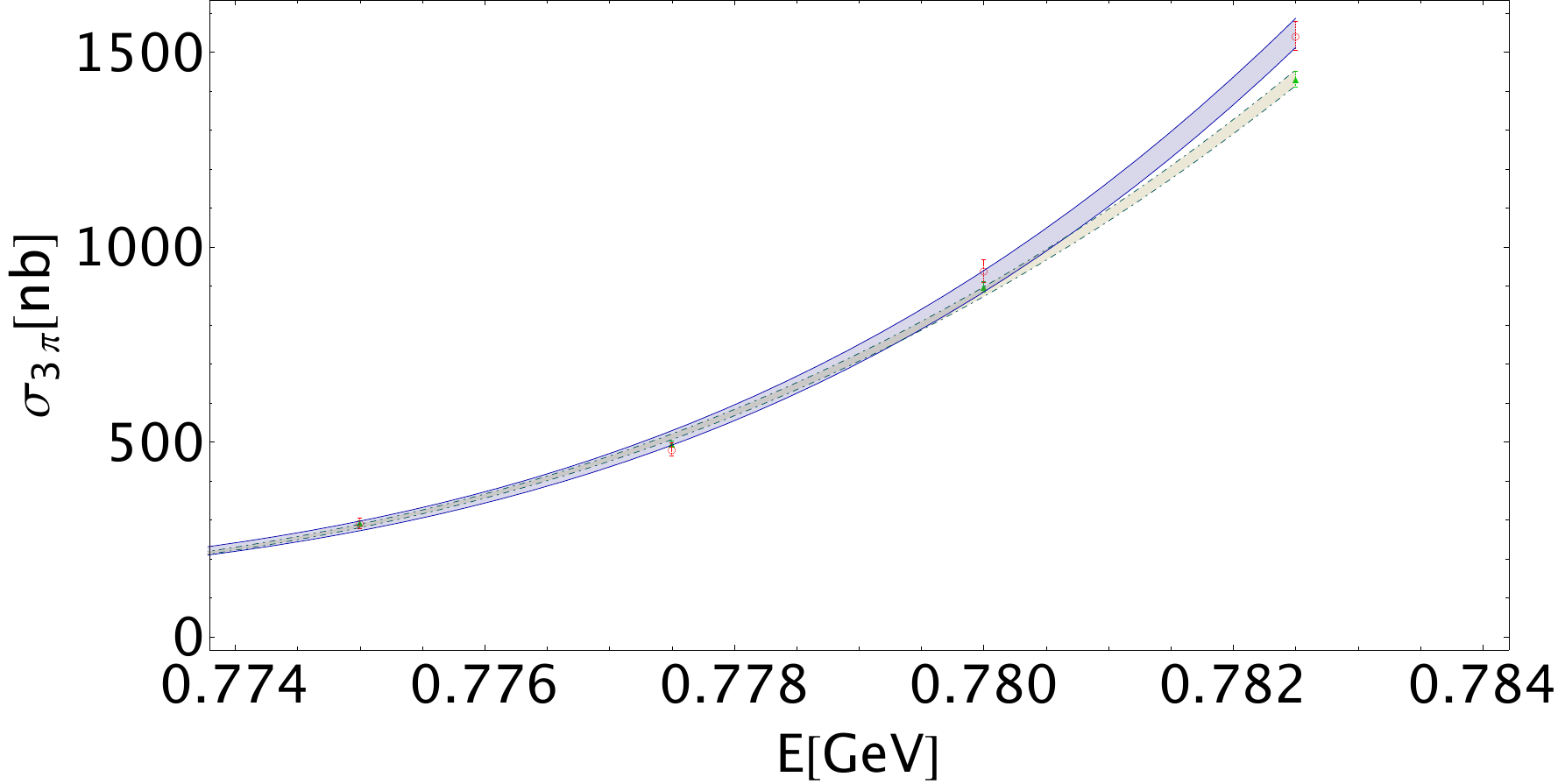}
\caption{\footnotesize  Fit of the BABAR (green triangle) and BELLE II (red circle) data in the region 0.63 to 0.783 GeV. } \label{fig:3pia}
\end{center}
\vspace*{-0.5cm}
\end{figure}

\begin{figure}[hbt]
\begin{center}
\includegraphics[width=7.5cm]{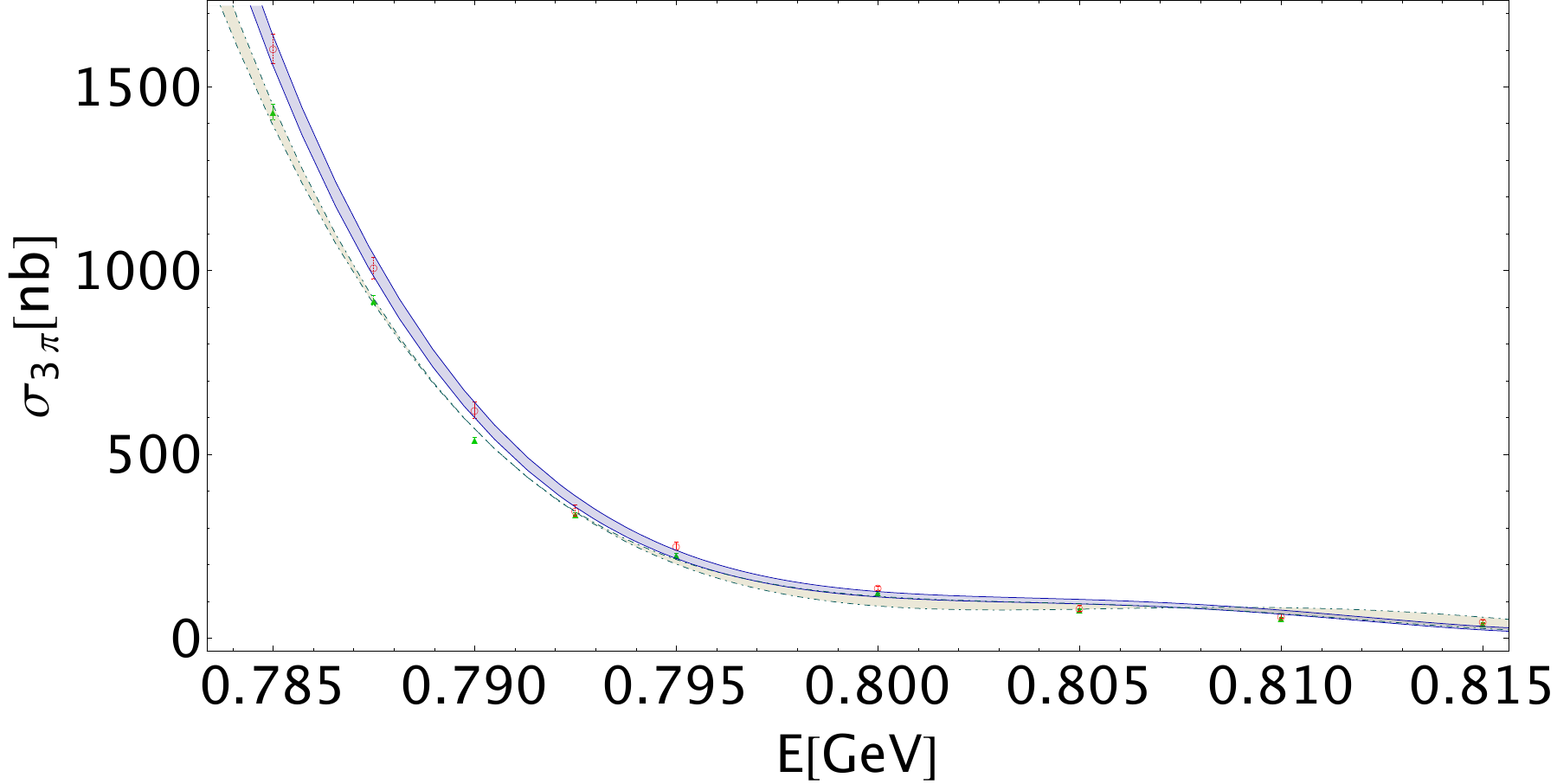}
\includegraphics[width=7cm]{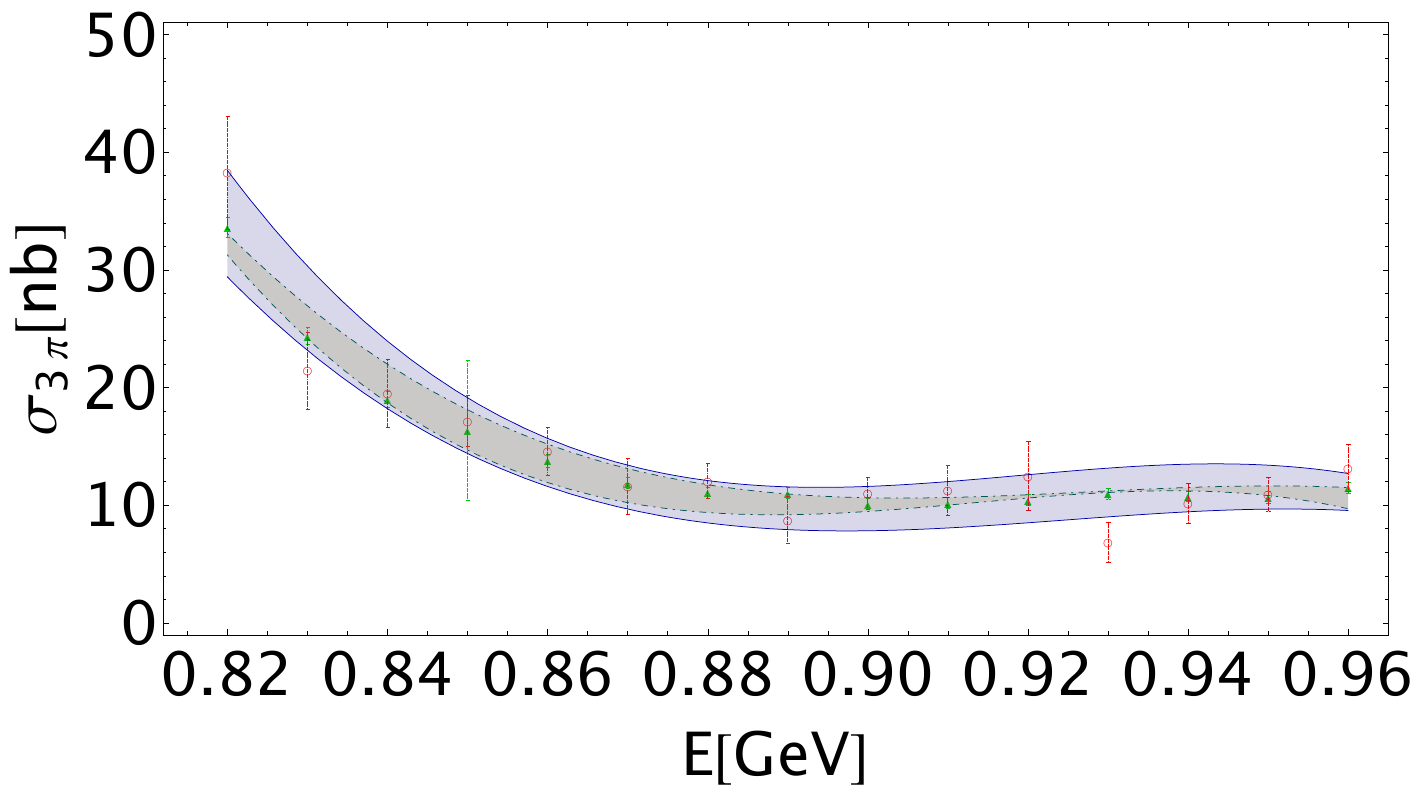}
\caption{\footnotesize  The same as in Fig.\,\ref{fig:3pia} but in the region 0.783 to 0.96 GeV. } \label{fig:3pib}
\end{center}
\vspace*{-0.5cm}
\end{figure}

Instead of the Narrow Width Approximation (NWA) used in Ref.\,\cite{SNe}, we use the $e^+e^-\to3\pi$ data from BABAR\,\cite{BABAR} (most accurate) and BELLE II\,\cite{BELLE} (most recent). We use as in the previous case the  optimized Mathematica FindFit package for fitting the data using some polynomials. The analysis is shown in Figs.\ref{fig:3pia} and \ref{fig:3pib}.  

We fit the data until 0.96 GeV where it starts to increase due to the presence of the $\phi$ meson and to the opening of some other channels. We deduce from the analysis\,\footnote{We extrapolate the fit until 0.99 GeV from which we shall use the full data of $R{e^+e^-}$ compiled by PDG\,\cite{PDG}. The change from 0.96 to 0.99 GeV increases the contribution to $a_\mu$ by $4\times 10^{-11}$.}\,:
\bea
a_\mu\vert_{lo}^{hvp}(3m_\pi\rar 0.99) &=& (333.7\pm 9.8)\times 10^{-11}
~~~~~{\rm BELLE II} , \nnb\\
&=& (313.3\pm 3.8)\times 10^{-11}~~~~~{\rm BABAR} 
\eea
where the BABAR data leads to a result 3 times more accurate  than the one of  BELLE II.  
In this channel,  about 43\% of the contribution comes from the region below the $\omega$-meson mass.
If instead, we use a NWA as in Ref.\,\cite{SNe}, one uses\,:
\beq
a_\mu^{\omega}\vert^{hvp}_{l.o}= \frac{3}{\pi}\frac{\Gamma^{ee}_\omega}{M_\omega} K_\mu(M_\omega^2).
\eeq
which leads to\,\cite{SNe}:
\beq
a_\mu\vert_{lo}^{hvp} \vert^{\omega}_{nwa}= (417\pm 14)\times 10^{-11}.
\eeq
The relative large error is due to the inaccurate value of the $\omega\to e^+e^-$ decay width.  The larger value of $a_\mu$ from the NWA than the one  from the Breit-Wigner cutted at 0.96 GeV may be due to the fact  that the NWA takes into account the effect of the tail of the $\omega$ width.  

We make a tentative average of the three results by multiplying the BABAR error by a conservative factor 2. In this way, we obtain the mean:
\beq
a_\mu\vert_{lo}^{hvp}(3m_\pi\rar 0.99) = (335.9\pm 5.5)\times 10^{-11},
\label{eq:amu-3pi}
\eeq
which we consider as a final estimate quoted in Table\,\ref{tab:amu1}.


The values obtained by BABAR $(459\pm 6)\times 10^{-11}$  and BELLE II $(481\pm 11)\times 10^{-11}$ are expected to be larger as they include all $3\pi$ contributions below 1.8 GeV.
\section{The $\phi(1020)$ $I=0$ isoscalar meson}
We estimate its contribution using a NWA. In this way, we obtain\,\cite{SNe}:
\beq
a_\mu\vert_{lo}^{hvp} \vert^{\phi}_{nwa}= (389.6\pm 4.6)\times 10^{-11}.
\eeq
Once we have isolated the $\phi$-meson contributions, we estimate the contributions from 0.96 to 1.875 GeV using the data compiled by PDG on the ratio $R^{ee}$.  The analysis has been done in ref.\,\cite{SNe}. 

\section{Light $I=0$ Isoscalar mesons from 0.96 to 1.875 GeV}
\subsection*{\hspace*{0.5cm} \b Data from 0.99 to 1.5 GeV}
We consider the $I=0$ contribution  substracted  in the $I=1$ analysis of the previous sections by using  the $SU(3)$ relation. We fit the data using a polynomial. We obtain\,\cite{SNe}\,: 
\beq
a_\mu\vert_{lo}^{hvp}(0.99\rar 1.5) = (44.3\pm 0.8)\times 10^{-11},
\eeq
\subsection*{\hspace*{0.5cm} \b Data from 1.5 to 1.875 GeV}
We consider the contributions of the $\omega(1650)$ and $\phi(1680)$. We use the parameters deduced from PDG22\,\cite{PDG}:
\bea
\omega(1650)&:&  M_\omega=(1670\pm 30)\,{\rm MeV},\,\,\,\,\,\,        \,\,\,\,\,  \,\,\,\Gamma^{ee}_\omega=1.35\pm 0.14\,{\rm keV},\,\,\,\,\,\,  \,\,\,\,\,\,  \,\,\Gamma^{h}_\omega=315\pm 35\,{\rm MeV} \nnb\\
\phi(1680)&:&  M_\phi=(1680\pm 20)\,{\rm MeV},\,\,\,\,\,\,   \,\,\,     \,\,\,\,\,  \,\,\,\Gamma^{ee}_\phi=0.18\pm 0.06\,{\rm keV},\,\,\,\,\,\,  \,\,\,\,\,\,  \,\,\Gamma^{h}_\psi=150\pm 50\,{\rm MeV}.
\eea

We use a Breit-Wigner for describing the $\omega(1650)$ and $\phi(1680)$ meson. We obtain\,:
\bea
a_\mu\vert_{lo}^{hvp}[\omega (1650)] &=& (24.3\pm 0.1)\times 10^{-11},\nnb\\
a_\mu\vert_{lo}^{hvp}[\phi (1680)] &=& (1.8\pm 0.9)\times 10^{-11},
\eea
\section{Light $I=0$ isoscalar mesons  contributions from $3m_\pi$ to  1.875 GeV}
Adding the sum of previous $I=0$ isoscalar mesons contributions, we deduce:
\beq
a_\mu\vert_{lo}^{hvp}(3m_\pi\rar 1.875) = (795.9\pm 7.7)\times 10^{-11},
\eeq
compiled in Table\,\ref{tab:amu1}. 

\section{Light $I=1\oplus 0$ light mesons from 1.875 to 3.68 GeV}
\subsection*{\hspace*{0.5cm} \b Data from $1.875\rar 2$ GeV}
\begin{figure}[hbt]
\begin{center}
\includegraphics[width=7.2cm]{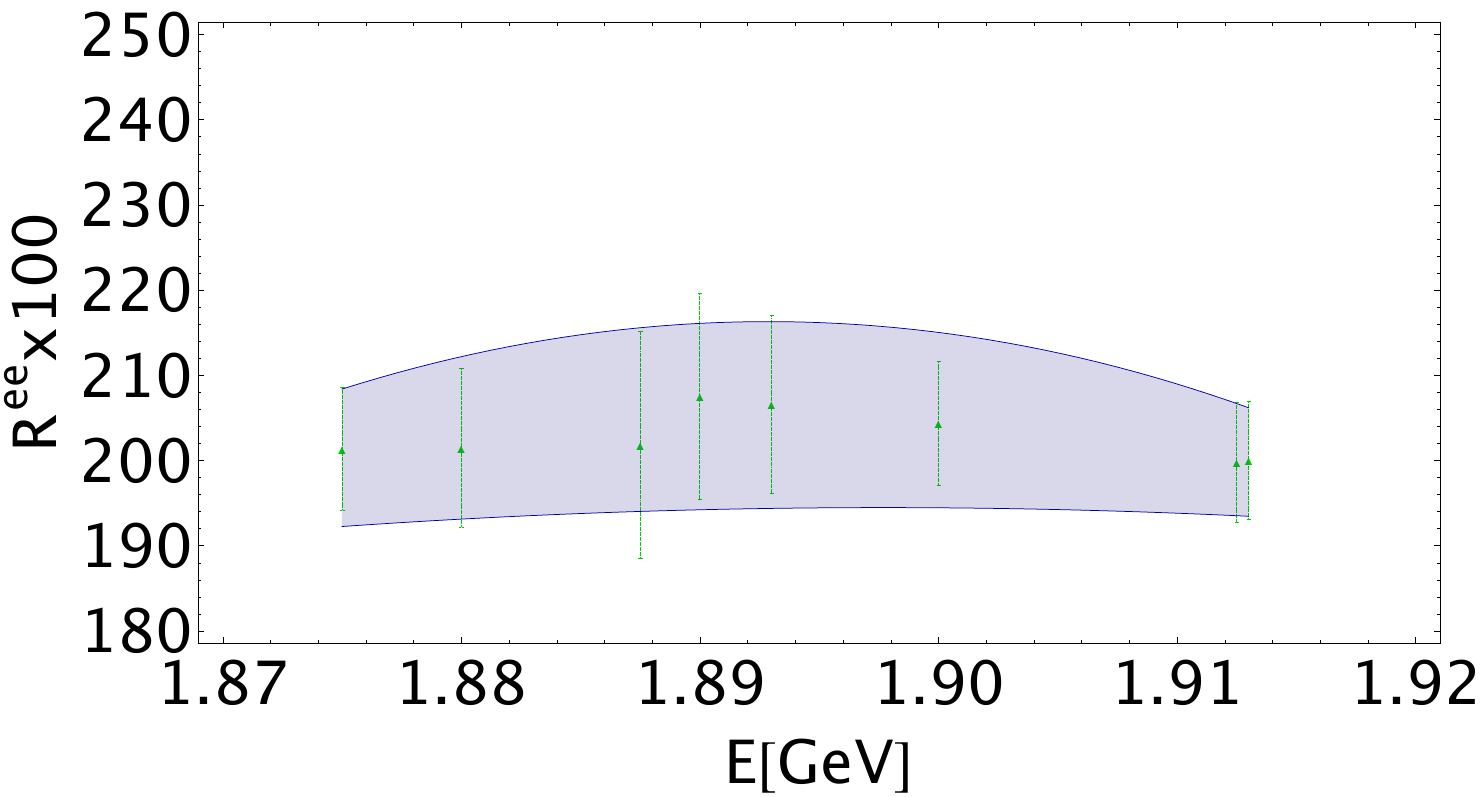}
\includegraphics[width=7.cm]{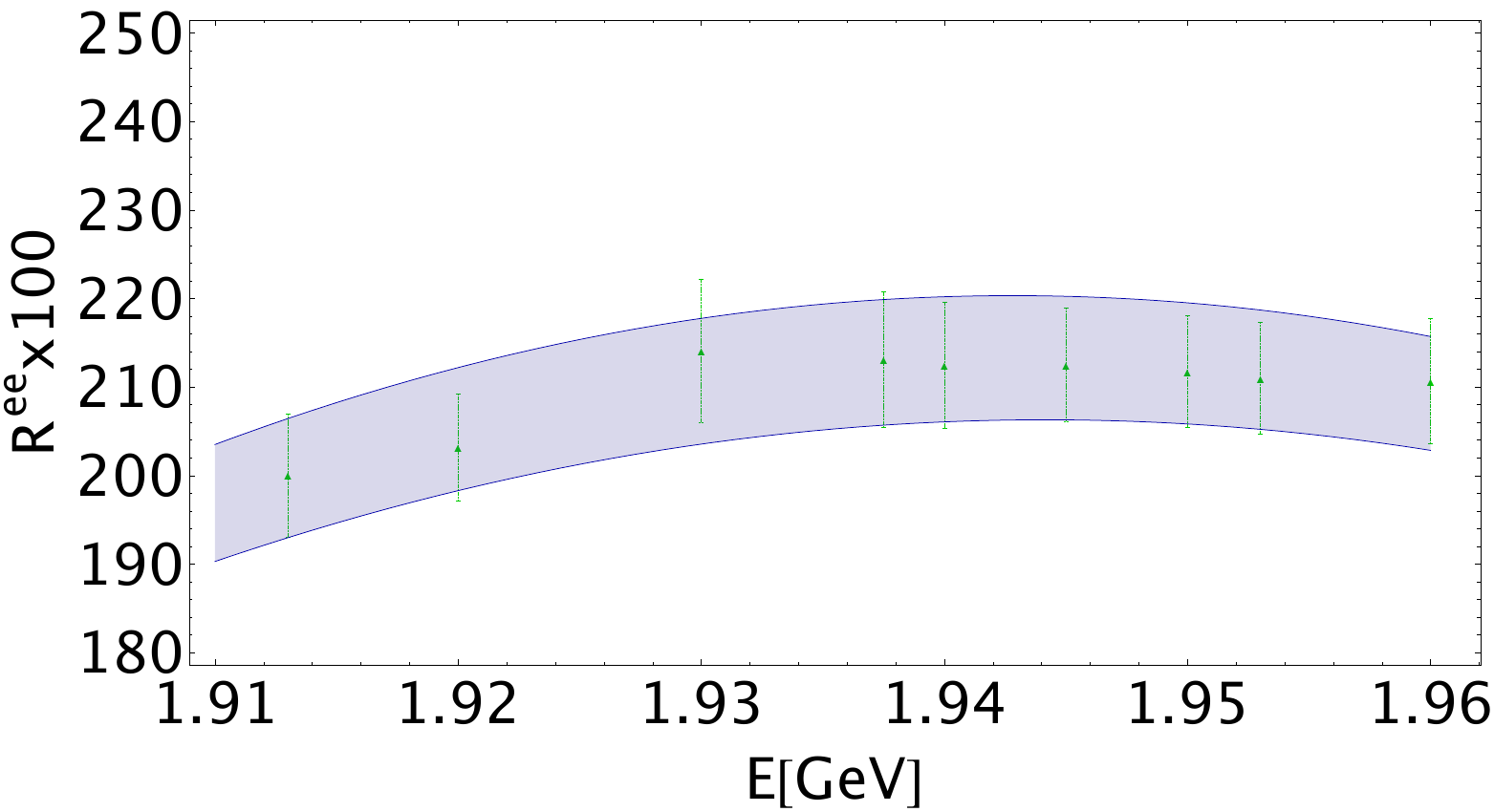}
\includegraphics[width=8cm]{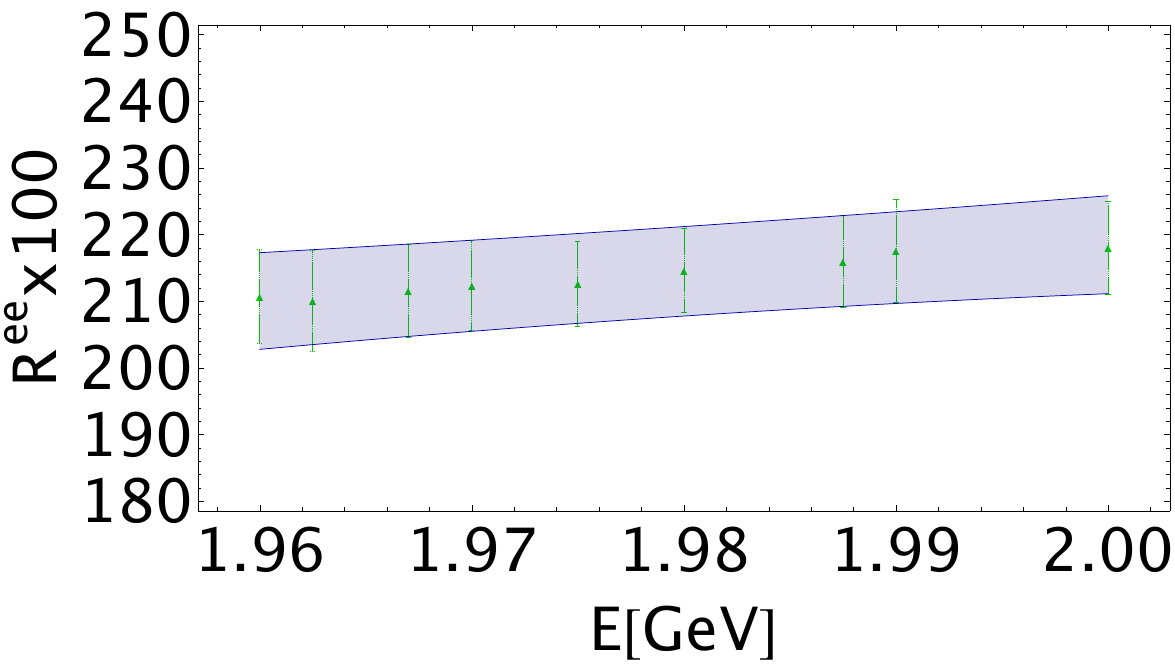}
\caption{\footnotesize  Fit of the PDG data in the region 1.875 to 2 GeV. } \label{fig:fit-R5}
\end{center}
\vspace*{-0.5cm}
\end{figure}

To get this contribution, we divide the data into three regions and fit with polynomials using the optimized $\chi^2$ Mathematica program FindFit. The fits of the data are shown in Fig.\ref{fig:fit-R5}. 
We obtain (see details in Table\,\ref{tab:amu1})\,::
\beq
a_\mu\vert_{lo}^{hvp}(1.875\rar 2) = (46.6\pm 1.0)\times 10^{-11}.
\eeq
\subsection*{\hspace*{0.5cm} \b QCD continuum from $2\rar 3.68$ GeV}
\begin{figure}[hbt]
\begin{center}

\includegraphics[width=12cm]{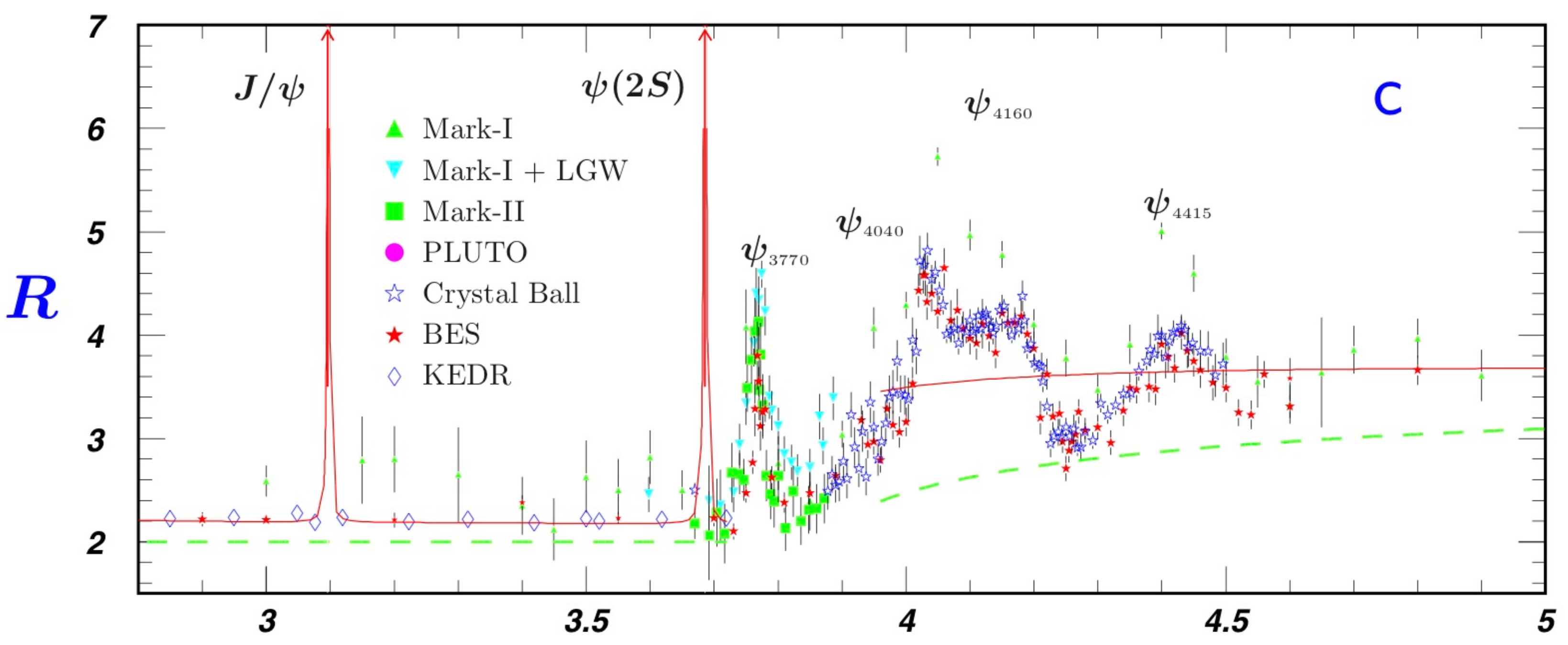}
\caption{\footnotesize  $e^+e^-\to$ Hadrons PDG data in the charmonium region from 2 to 5 GeV from PDG\,\cite{PDG}.  The broken green line is the naive parton model prediction. The continuous red line is the QCD prediction including 3-loop PT corrections.} \label{fig:charm}
\end{center}
\vspace*{-0.5cm}
\end{figure}

The data in this region are well fitted by the QCD expression of the spectral function for 3 flavours as one can see in Fig.\,\ref{fig:charm} from PDG\,\cite{PDG}. 

To the massless PT expression known to order $\alpha_s^4$, we  add the contributions of the quark and gluon condensates of dimensions $D=4,6$.  We add the quadratic  $m^2_s$-corrections to order $\alpha_s^3$ and the quartic $m_s^4$-mass corrections to order $\alpha_s^2$ (see\,\cite{SNe} and references quoted therein). We obtain:
\beq
a_\mu\vert_{lo}^{hvp}(2\rar 3.68) = (247.2\pm 0.3)\times 10^{-11},
\eeq

\section{Light $I=1\oplus 0$ light mesons from $2m_\pi$ to 3.68 GeV}
Adding the different contributions of the $I=0$ and 1 light mesons below 2 GeV $\oplus$ the 
corresponding QCD continuum from 2 to 3.68 GeV, we obtain:
\beq
a_\mu\vert_{lo}^{hvp}(2m_\pi\rar 3.68) = (6791.0\pm 37.1)\times 10^{-11},
\eeq

\section{Heavy quarkonia contributions}

\subsection*{\hspace*{0.5cm} \b Charmonium from $J/\psi$ to 10.5 GeV}
\subsection*{\hspace*{1.0cm} \d Narrow resonances}
We estimate the contributions of $\psi(1S),  \psi(2S)$ and $\psi(3773)$ using a NWA. 
 The result is given in Table\,\ref{tab:amu1}.
\begin{figure}[H]
\begin{center}
\includegraphics[width=6.cm]{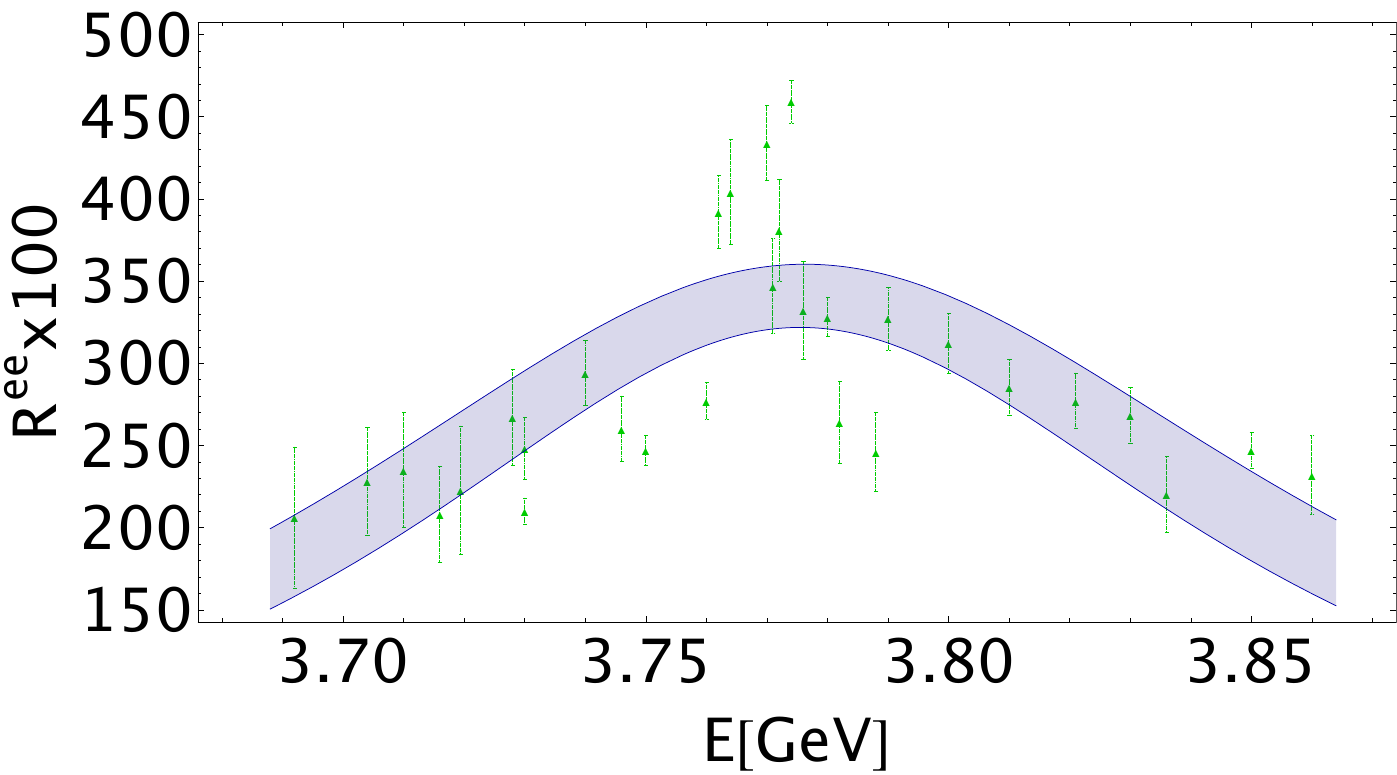}
\includegraphics[width=6.cm]{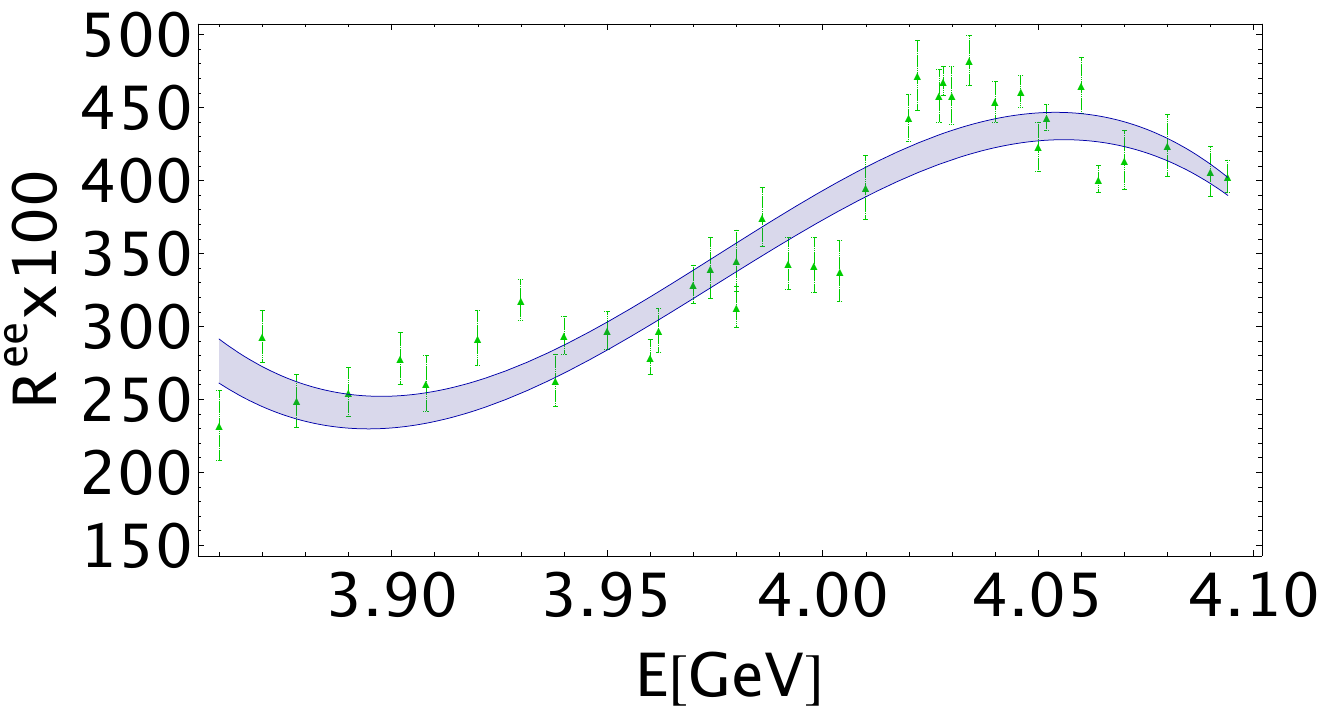}\\
\includegraphics[width=6.cm]{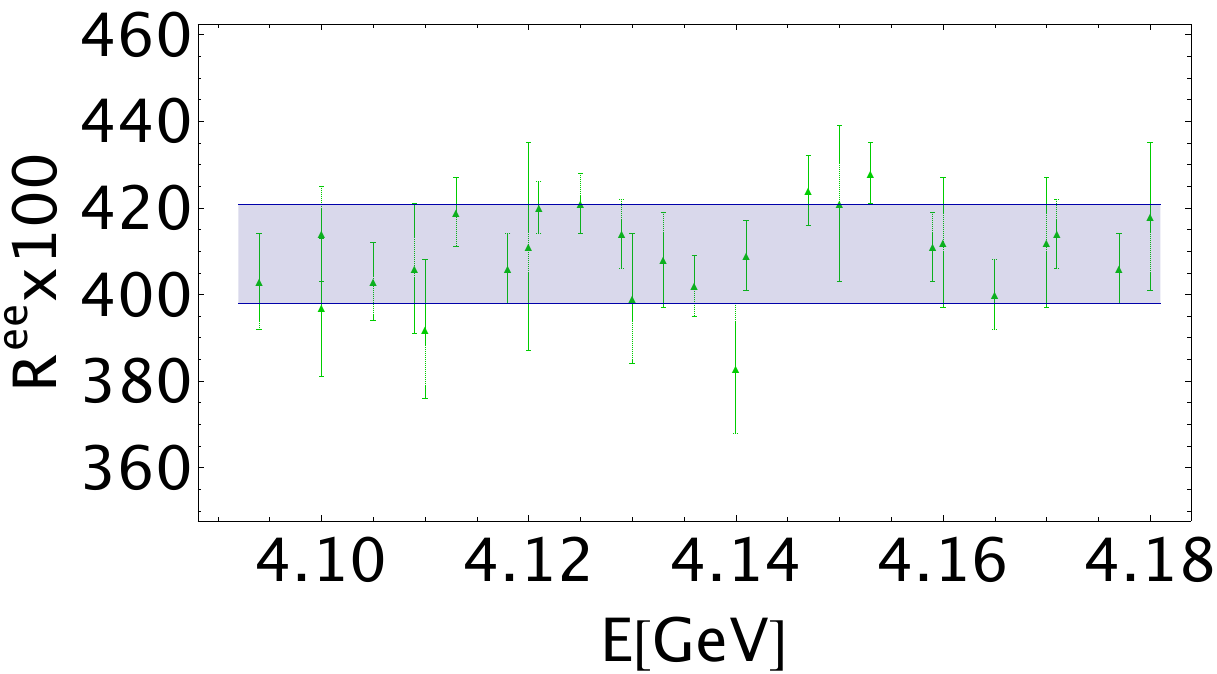}
\includegraphics[width=6.cm]{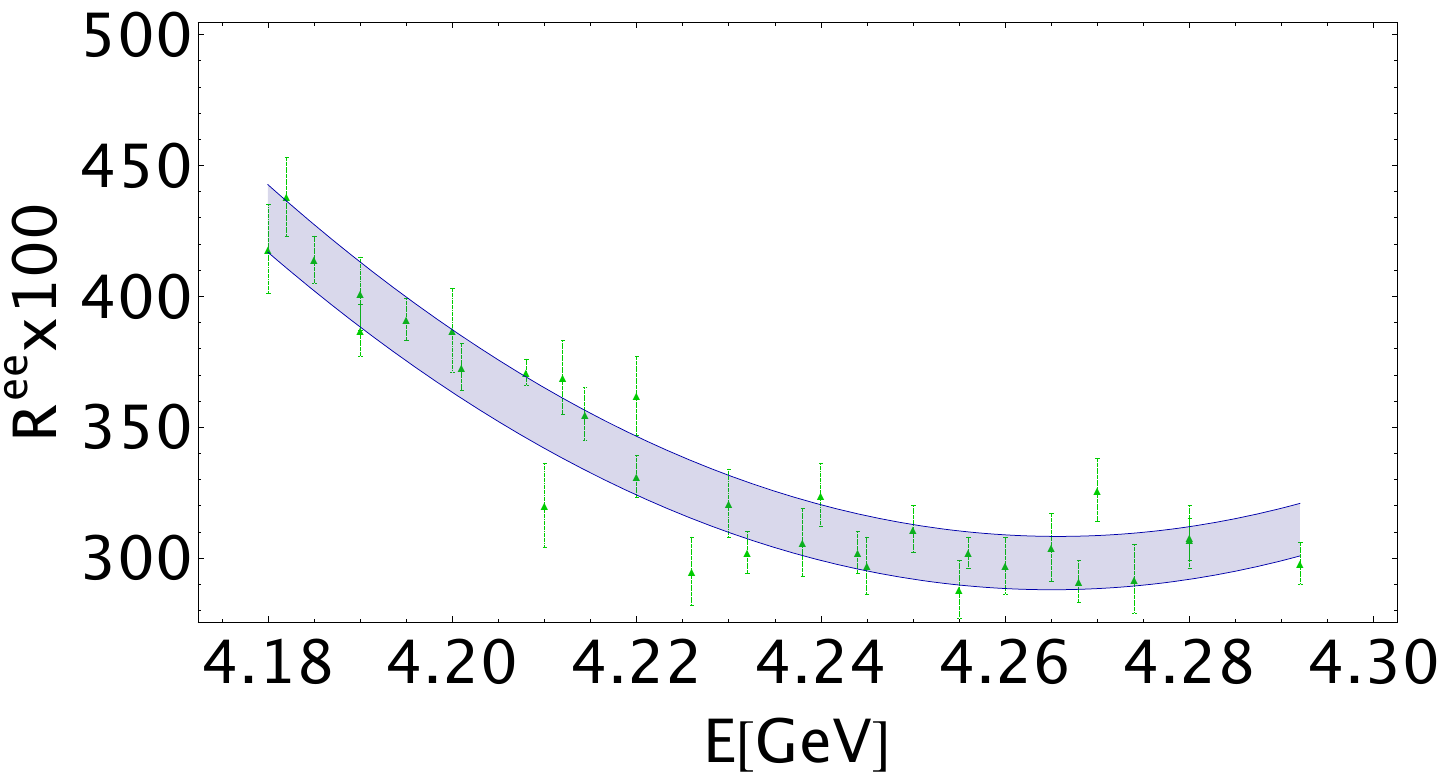}\\
\includegraphics[width=7cm]{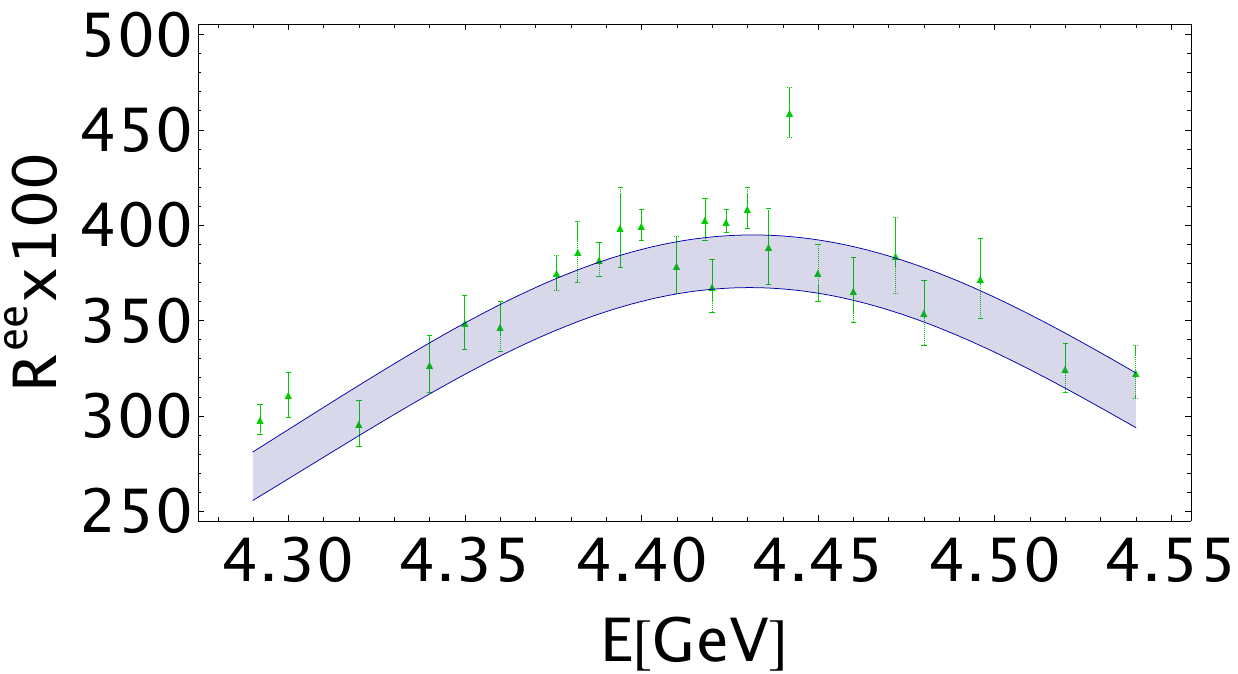}
\caption{\footnotesize  Fits of the $e^+e^-\to$ Hadrons data in the charmonium region from 3.68 to 4.55 GeV from PDG\,\cite{PDG}. } \label{fig:fit-R7}
\end{center}
\vspace*{-0.5cm}
\end{figure} 
\begin{figure}[H]
\begin{center}
\includegraphics[width=12cm]{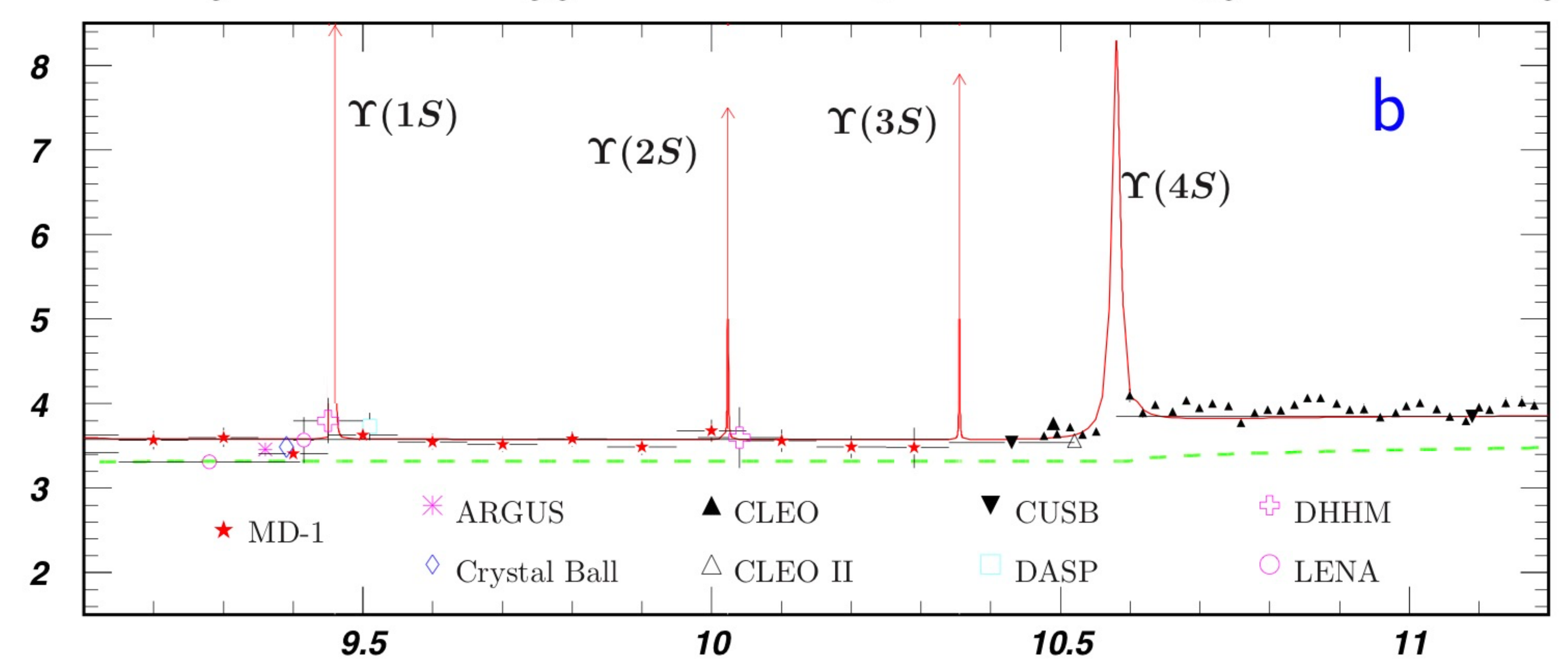}
\caption{\footnotesize  $e^+e^-\to$ Hadrons data in the bottomium region from 9 to 11.5 GeV from PDG\,\cite{PDG}. The broken green line is the naive parton model prediction. The continuous red line is the QCD prediction including 3-loop PT corrections. } \label{fig:bottom}
\end{center}
\vspace*{-0.5cm}
\end{figure} 

\subsection*{\hspace*{1.0cm} \d Region from 3.68 to 4.55 GeV}
We divide this region into 5 subregions as shown in Fig.\,\ref{fig:fit-R7}. The region from 3.68 to 3.86 GeV is better fitted using a Breit-Wigner while, for the others,  we use polynomials. The different contributions are shown in Table\,\ref{tab:amu1}. 
\subsection*{\hspace*{0.5cm}\d Region from 4.55 to 10.50 GeV }
As shown in Figs.\,\ref{fig:charm} and \ref{fig:bottom}, the data without resonance peaks are well fitted by QCD for 4 flavours. We add to the previous QCD expressions the charm contributions with $m_c^2$ and $m_c^4$  mass corrections where $\overline{m}_c(\overline{m}_c)=$ 1266(6) MeV\,\cite{SNREV1}. The contribution is given in Table\,\ref{tab:amu1}. 
\subsection*{\b   $a_\mu\vert^{hvp}_{l.o}$ from bottomium}
\subsection*{ \hspace*{0.5cm} \d $\Upsilon(1S\to 11.02)$ contributions}
We use a NWA to estimate the contributions of $\Upsilon(1S), \Upsilon(2S), \Upsilon(3S)$ and $\Upsilon(4S)$. 
The results are reported in Table\,\ref{tab:amu1}. 
\subsection*{ \hspace*{0.5cm} \d QCD continuum contribution from  10.59 GeV to $2m_t$}
We add the $b$-quark contribution to the previous QCD expression where $b$-quark mass corrections to order $a_s^3\overline{m}_b^2/t$ and $a_s^2\overline{m}_b^4/t^2$ are included. We use\,\: $\overline{m}_b(\overline{m}_b)=$ 4202(8) MeV\,\cite{SNREV1}.
For the analysis, we consider the region from 10.59 MeV to $2m_t$ just after the $\Upsilon(4S)$ where the QCD continuum is expected to smear the  $\Upsilon(10860, 11020)$  and some eventual  higher resonances. The result is given in Table\,\ref{tab:amu1}. 

   {\footnotesize
\begin{table}[H]
\vspace*{-1.cm} 
\setlength{\tabcolsep}{1.pc}
  \begin{center}
    { \footnotesize
  \begin{tabular}{| llll |}
\hline
\rowcolor{yellow}\boldmath$\sqrt{t}$\,\bf [GeV] &\boldmath$a_\mu\vert^{hvp}_{l.o}\times 10^{11} $&\boldmath$a_\tau\vert^{hvp}_{l.o}\times 10^{9} $&$\Delta\alpha^{(5)}_{had}(M_Z^2)\times 10^5$\\ 
 \hline 

{\bf Light I=1}&&&\\
$\rho(2m_\pi\to 0.60)$&$1069.0\pm 22.1$&$85.1\pm 0.6$&$12.38\pm 0.08$\\
$\rho(0.60\to 0.778)$&$2795.9\pm 26.0$&$943.7\pm10.3$&$182.87\pm 1.97$ \\
$\rho(0.778\to 0.993)$&$1244.4\pm 8.6$&$1797.3\pm 20.3$&$117.82\pm 0.98$\\
$0.993\to 1.5$&$354.4\pm 6.7$&$228.2\pm 4.1$&$67.3\pm 1.14$\\
$1.5\to 1.875$&$237.6\pm 5.7$&$206.8\pm 4.9$&$80.11\pm 1.9$\\

{\it Total Light I=1 ($\leq 1.875$)}&$\it 5701.3\pm 36.3$&$\it 2453.3\pm 11.9$&$\it 483.3\pm 4.4$\\
{\bf Light I=0}&&&\\
$3\pi: ~3m_\pi\to 0.99$&$335.9\pm 5.5$&$163.3\pm 5.3$&$33.6\pm 1.1$\\
$\phi$ (NWA) &$389.6\pm 4.6$&$20.6\pm 0.2$&$51.2\pm 0.6$\\
$0.993\to 1.5$ & $44.3\pm 0.8$&$28.5\pm 0.5$&$8.4\pm 0.1$\\
$\omega(1650)$(BW)&$24.3\pm 0.1$&$16.7\pm 0.1$&$5.2\pm 0.1$ \\
$\phi(1680)$(BW)&$1.8\pm 0.9$&$1.3\pm 0.6$&$0.4\pm 0.2$ \\
{\it Total Light I=0 ($\leq 1.875$)}&$\it 795.9\pm 7.3$&$\it 230.4\pm 5.3$&$\it 99.4\pm 1.3$\\
{\bf Light I=\boldmath $0\, \oplus \,1$}&&&\\
$1.875\to 1.913$ & $14.7\pm 0.7$&$14.4\pm 0.7$&$6.3\pm 0.3$\\
$1.913\to 1.96$ & $17.6\pm 0.6$&$17.5\pm 0.6$&$7.9\pm 0.3$\\
$1.96\to 2$ & $14.3\pm 0.5$&$14.5\pm 0.5$&$6.7\pm 0.2$\\
$2\to 3.68$:  QCD $(u,d,s)$ & $247.2\pm 0.3$&$308.3\pm 0.5$&$202.8\pm 0.5$\\
{\it Total Light I=$0\, \oplus \,1$ ($1.875\to 3.68$)}&$\it 293.8\pm 1.1$&$\it 354.7\pm 1.2$&$\it 223.7\pm 0.7$\\
\rowcolor{greenli}{ Total Light I=$0\, \oplus \,1$ ($2m_\pi\to3.68$)}&$ 6791.0\pm 37.1$&$3038.4\pm 24.5$&$806.4\pm 4.6$\\
{\bf Charmonium} &&&\\
$J/\psi (1S)$ (NWA) &$65.1\pm 1.2$&$92.7\pm 1.8$&$73.5\pm 1.4$\\
$\psi (2S)$ (NWA) &$16.4\pm 0.6$&$26.0\pm 0.9$&$26.1\pm 0.8$\\
$\psi (3773)$ (NWA) &$1.7\pm 0.1$&$2.7\pm 0.2$&$2.9\pm 0.2$\\
{\it Total $J/\psi (NWA)$} &$\it 83.2\pm 1.4$&$\it 121.4\pm 2.0$&$\it 102.5\pm 1.6$\\
$3.69\to 3.86$ & $11.4\pm 1.0$&$18.3\pm 1.6$&$19.0\pm 1.6$\\
$3.86\to4.094$ & $16.6\pm 0.5$&$27.5\pm 0.8$&$30.9\pm 0.9$\\
$4.094\to4.18$ & $6.6\pm 0.2$&$11.2\pm 0.3$&$13.2\pm 0.4$\\
$4.18\to4.292$ & $6.5\pm 0.2$&$11.2\pm 0.4$&$13.7\pm 0.5$\\
$4.292\to4.54$ & $11.8\pm 0.6$&$20.7\pm 0.8$&$26.8\pm 1.1$\\
$4.54\to10.50$:  QCD $(u,d,s,c)$ & $92.0\pm 0.0$&$186.2\pm 0.0$&$458.7.3\pm 0.1$\\
{\it Total Charmonium ($3.69\to 10.50$)}&$\it 145.1\pm 1.3$&$\it 275.6\pm 2.0$&$\it 564.9\pm 2.2$\\
\rowcolor{greenli}{ Total Charmonium }&$ 228.1\pm 1.9$&$396.5\pm 2.8$&$664.8\pm 2.7$\\
 {\bf Bottomium} &&&\\
$\Upsilon (1S)$ (NWA) &$0.54\pm 0.02$&$1.25\pm 0.07$&$5.65\pm 0.29$\\
$\Upsilon(2S)$ (NWA) &$0.22\pm 0.02$&$0.51\pm 0.06$&$2.54\pm 0.29$\\
$\Upsilon (3S)$ (NWA) &$0.14\pm 0.02$&$0.33\pm 0.04$&$1.77\pm 0.23$\\
$\Upsilon(4S)$ (NWA) &$0.10\pm 0.01$&$0.23\pm 0.03$&$1.26\pm 0.16$\\
$\Upsilon(10.86\,\oplus\,11)$ (NWA) &$0.1\pm 0.06$&$0.20\pm 0.06$&$1.67\pm 0.39$\\
{\it Total Bottomium (NWA)}&$\it1.0\pm 0.1$&$\it 2.3\pm 0.1$&$\it 11.2\pm 0.5$ \\
$Z-pole $&-&-&29.2\,\cite{YND17}\\
$10.59\to2m_t$\,:  QCD $(u,d,s,c,b)$ & $22.4\pm 0.3$&$57.5\pm 0.1$&$1282.9\pm 1.2$\\
\rowcolor{greenli}{Total Bottomium }&$ 23.4\pm 0.3$&$59.8\pm 0.1$&$1323.3\pm1.3$\\
$2m_t\to\infty$\,: QCD $(u,d,s,c,b,t)$&0.03&0.08&-28.2\\
\hline
\rowcolor{yellow}\bf Total sum &\boldmath$7042.5\pm 37.2$&\boldmath$3494.8\pm 24.7$&\boldmath$2766.3\pm 4.5$\\

   \hline
 
\end{tabular}}
 \caption{ 
  $a_\mu\vert^{hvp}_{l.o}$, $a_\tau\vert^{hvp}_{l.o}$  and $\Delta \alpha(M^2_z)$ within our parametrization of the compiled PDG\,\cite{PDG} $\oplus$ the recent CMD-3\,\cite{CMD3}. The contributions of $\pi^0\gamma$ and $\eta\gamma$ are not included inthis Table. We have not updated the estimate of $a_\tau\vert^{hvp}_{l.o}$ and $\Delta \alpha(M^2_z)$ as the effect of the new fit is almost negligible.}\label{tab:amu1} 
 \end{center}
\end{table}
} 
\subsection*{ \hspace*{0.5cm} \d QCD continuum contribution from  $2m_t\to\infty$}
Due to the heaviness of the top quark mass, we shall use the approximate Schwinger formula near the $\bar tt$ threshold
for a much better description of the spectral function up to order $\alpha_s$ due to the top quark:
\beq
R_t^{ee}=\frac{4}{3} v\frac{(3-v^2)}{2}\Big{[} 1+\frac{4}{3}\alpha_sf(v)\Big{]},
\eeq
with:
\beq
f(v)=\frac{\pi}{2\, v}-\frac{(3+v)}{4}\ga \frac{\pi}{2}-\frac{3}{4\pi}\dr\,:\,\,\,\,\,\, v=\ga 1-\frac{m_t^2}{t}\dr^{1/2}.
\eeq
Here $m_t$ is the on-shell top quark mass which we fix to be\,\cite{PDG}\,\footnote{We should note that the definition of the top quark mass from different experiments is still ambiguous.}: 
\beq
\overline{m}_t(\overline{m}_t)=(172.7\pm 0.3)\,{\rm  GeV},
\label{eq:mt}
\eeq
 from some  direct measurements.
We add to this expression the  one due to $\alpha_s^2$ and $\alpha_s^3$, given in Ref.\,\cite{SNe}, within the $\overline {MS}$-scheme.  For this $\overline {MS}$-scheme expression  in terms of running mass, we need the value of the RGI top mass:
\beq
\hat m_t=(254\pm 0.4)\,{\rm GeV},
\eeq
deduced from Eq.\,\ref{eq:mt} (for reviews, see e.g.\,\cite{SNB1,SNB2}).
Adding the above expressions to the ones in the previous sections, we obtain the result given in Table\,\ref{tab:amu1}. 

\section{Results and Comments  for $a_\mu\vert^{hvp}_{lo}$}
\subsection*{ \hspace*{0.5cm} \b Final results for $a_\mu\vert^{hvp}_{lo}$ and comparison with Ref.\,\cite{SNe}}
\subsection*{ \hspace*{0.5cm} \d  Results without $e^+e^-\to \pi^0\gamma+\eta\gamma$}

In the present paper, we have fitted only the CMD-3 data below 0.993 GeV while in Ref.\cite{SNe}, we have
combined in some sub-regions  the data from PDG22 and CMD-3. In this way the contribution from the $2\pi$ has slightly increased in the present work.
However, this increase has been partially compensated by the one due to 3$\pi$ with which we have replaced the NWA  $\omega$-meson used in Ref.\,\cite{SNe}. The value of  $a_\mu\vert^{hvp}_{lo}$ is\,:
\bea
a_\mu\vert^{hvp}_{lo}&=&(7037\pm 39)\times 10^{-11} ~~~:~ \cite{SNe}  
 \nnb \\
&=&  (7042.5\pm 37.2)\times 10^{-11} ~~~:~{\rm this~work}. 
\label{eq:amu-res}
\eea
It is remarkable that the central value is about the same as the one  $(7020\pm 800)\times 10^{-11}$  in my PhD 3rd cycle thesis\,\cite{SN76,CALMET} in 1976 but the accuracy has been improved by a factor 22 thanks to the amount of experimental efforts done during  this period. This result also agrees with the one\,:  $(7020.6\pm 75.6)\times 10^{-11}$ obtained in Ref.\,\cite{SNamu} but 2 times more precise.
\subsection*{ \hspace*{0.5cm} \d Contribution of the  $e^+e^-\to \pi^0\gamma+\eta\gamma$ process.}
\begin{figure}[hbt]
\begin{center}
\includegraphics[width=6cm]{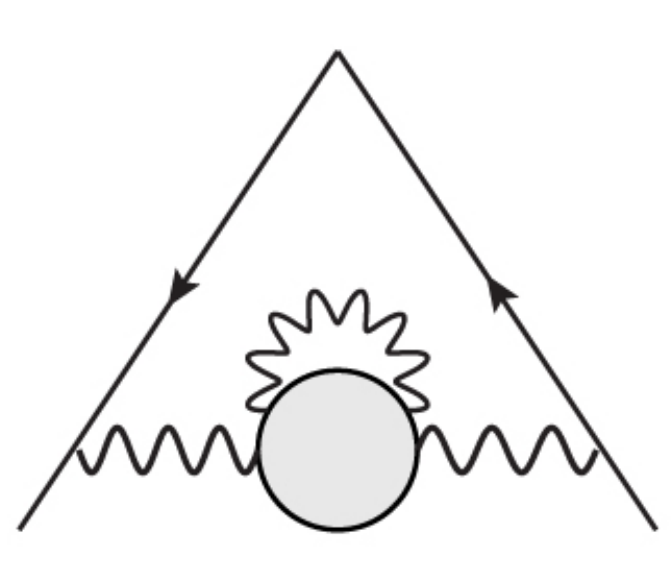}
\caption{\scriptsize Diagrammatic representation of $e^+e^-\to P\gamma (P\equiv \pi^0,\eta)$ contributions  to $a_\mu\vert^{hvp}_{lo}$ }\label{fig:amu-pgamma}
\end{center}
\vspace*{-0.5cm}
\end{figure}  
This contribution is shown in Fig.\,\ref{fig:amu-gamma}. 
We take the average of the estimates ($P\equiv \pi^0+\eta$)\,:
\beq
a_\mu^{P\gamma}\vert^{hvp}_{lo} = (51.6\pm 1.03)\times 10^{-11},\cite{NOMURA}\,\,{\rm and}
\,\, (50.6\pm 1.02)\times 10^{-11},\cite{DAVIER2},
\eeq
which leads to\,:
\beq
a_\mu^{P\gamma}\vert^{hvp}_{lo} = (51.05\pm 0.72)\times 10^{-11}.
\label{eq:pi0gamma}
\eeq
\subsection*{ \hspace*{0.5cm} \d  Final Results including $e^+e^-\to \pi^0\gamma+\eta\gamma$}
To get the final results,  we add Eq.\,\ref{eq:amu-res} and Eq.\,\ref{eq:pi0gamma}. We obtain\,:
\beq
a_\mu^{tot}\vert^{hvp}_{lo}=(7094\pm 37)\times 10^{-11}.  
\label{eq:amu-final}
\eeq
\subsection*{ \hspace*{0.5cm} \b Comparison with some other $e^+e^-$ data driven results}
Among others, the works of Refs.\,\cite{NOMURA}  and \cite{DAVIER2} using $e^+e^-\to$ Hadrons data are  mostly  quoted. A sharp comparison with these works cannot be easily done  as each authors choose different cuts when connecting different data.  They also estimate the contributions of some specific exclusive channels while we work with data of their sum or/and inclusive ones. Quoting the total sum of the results from different exclusive modes, they obtain\,:
\bea
a_\mu\vert^{hvp}_{lo} &=&(6928\pm 24)\times 10^{-11}~~~:~~\cite{NOMURA}
\nnb\\
&=&  (6940\pm 40)\times 10^{-11} ~~~:~~\cite{DAVIER2}. 
\eea
We notice that the two determinations are respectively lower by\,\footnote{A comparison of the different $e^+e^-\to$ Hadrons and $\tau$-decay data and Lattice results on the Adler function is discussed in Ref.\,\cite{PICH}.}\,:
\bea
\Delta a_\mu\vert^{hvp}_{lo}\,\cite{NOMURA}&\equiv& a_\mu\vert^{hvp}_{lo}\,[{\rm this~work}] - a_\mu\vert^{hvp}_{lo}\,\cite{NOMURA}
=(166\pm 31)\times 10^{-11}\nnb\\
\Delta a_\mu\vert^{hvp}_{lo}\,\cite{DAVIER2}&\equiv& a_\mu\vert^{hvp}_{lo}\,[{\rm this~work}] - a_\mu\vert^{hvp}_{lo}\,\cite{DAVIER2}= (154\pm 39)\times 10^{-11}. 
\label{eq:comparison}
\eea
\subsection*{ \hspace*{0.5cm} \d  $e^+e^-\to$ Hadrons  data  below 1.937 GeV\,\cite{NOMURA}}
-- To track the origin of the discrepancy, we compare the total sum of the contributions of the exclusive modes of the light meson channels up to 1.937 GeV from Ref.\,\cite{NOMURA} which are:
\beq
a_\mu\vert^{hvp}_{lo}(\leq 1.937) =(6393\pm 23)\times 10^{-11}
\eeq
to the sum of our estimate until 1.875 GeV given in Table\,\ref{tab:amu1}$\oplus  P\gamma$ process:
\beq
a_\mu\vert^{hvp}_{lo}(\leq 1.875) =(6497\pm 37\oplus 51\pm 0.7)\times 10^{-11}.
\label{eq:amu18}
\eeq
The difference between the two determinations is  $(155\pm 31)\times 10^{-11}$ which is about the same as in the total sum.  

-- By comparing our result up to 0.993 GeV with the one of \cite{NOMURA} taken until 1.937 GeV, we notice that at least $98\times 10^{-11}$ comes from the $2\pi$ channel which can be understood from the CMD-3 and PDG data plotted in Fig.\,\ref{fig:rho1} because in the regions from 0.4 to 0.6 GeV and 0.68 to 0.77 GeV, the CMD-3 data are systematically above the PDG22 compilation.

\subsection*{ \hspace*{0.5cm} \d $e^+e^-\to$ Hadrons data below $(1.8\sim 1.875)$  GeV \,\cite{DAVIER2}}
We track the difference between Ref.\,\cite{DAVIER2} and ours by examining two regions. 

-- We compare the results below $(1.8\sim 1.875)$ GeV:
\beq
a_\mu\vert^{hvp}_{lo}(\leq 1.8)\,\cite{DAVIER2} = (6356\pm 40)\times 10^{-11},
\eeq
which is $(192\pm 39)\times 10^{-11}$ lower than our result in Eq.\,\ref{eq:amu18}.

-- We also compare the results from 1.8 to 3.7 GeV:
\bea
a_\mu\vert^{hvp}_{lo}(1.8\to 3.7)\,\cite{DAVIER2} &=& (334.5\pm 7.1)\times 10^{-11}, \nnb\\
a_\mu\vert^{hvp}_{lo}(1.875\to 3.68)\,[\rm this~work] &=& (293.8\pm 1.1)\times 10^{-11}
\eea
where the result of \cite{DAVIER2} is $(41\pm 3)\times 10^{-11}$ higher than ours.

-- For the total sum of $a_\mu\vert^{hvp}_{lo}$ in all regions, the two differences tend to compensate and lead to the final underestimate of about $154\times 10^{-11}$ quoted in Eq.\,\ref{eq:comparison}.
\subsection*{ \hspace*{0.5cm} \b  Comparison of the $2\pi$ contributions from $e^+e^-$ and $\tau$-decays }
\subsection*{ \hspace*{0.5cm} \d  Our result from $e^+e^-$}
\begin{figure}[hbt]
\begin{center}
\includegraphics[width=6cm]{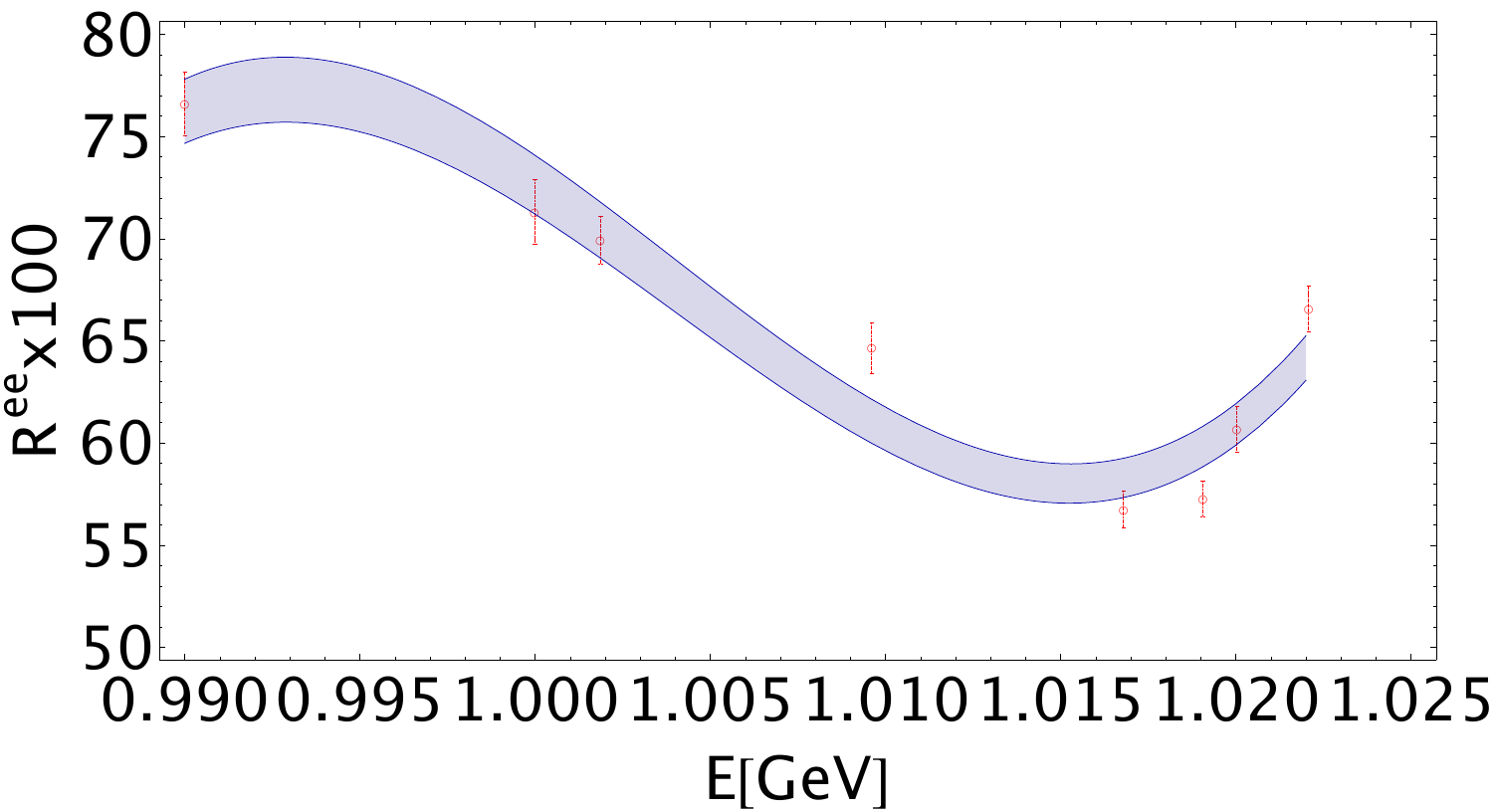}
\includegraphics[width=6cm]{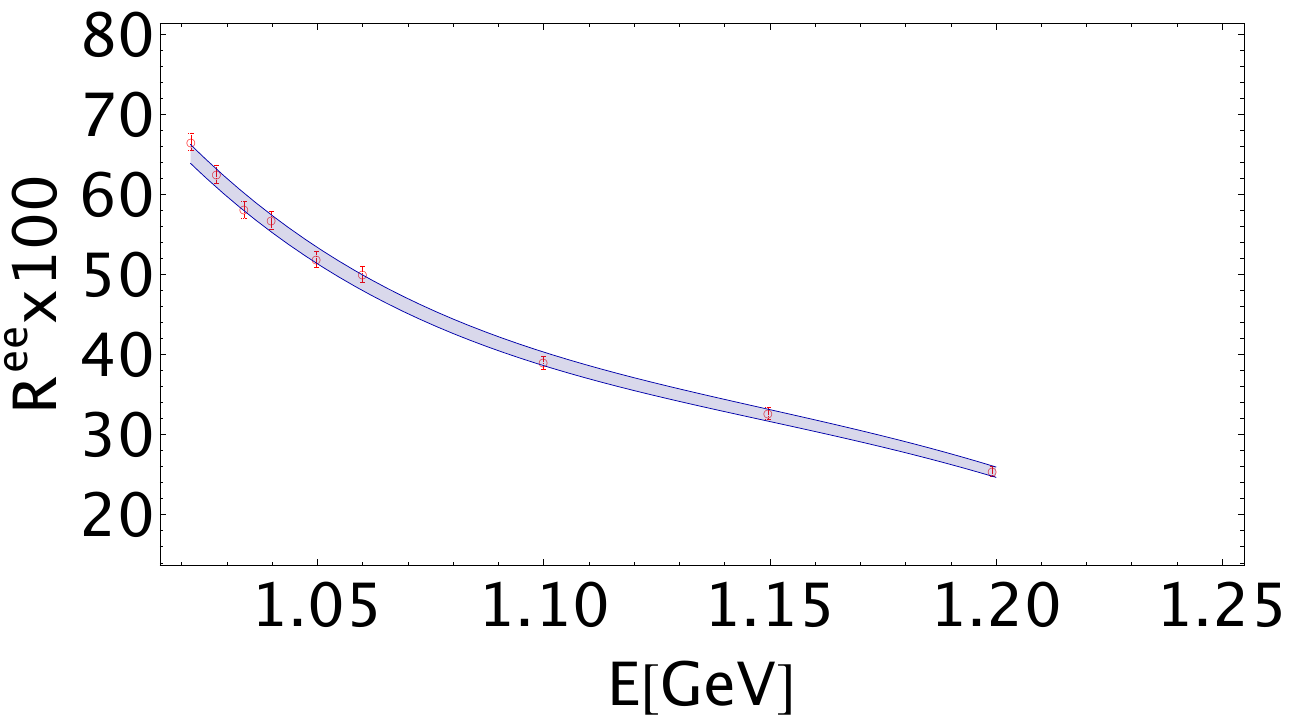}
\caption{\scriptsize Fit of the CMD-3 data from 0.99 to 1.2 GeV }\label{fig:cmd3-1.2}
\end{center}
\vspace*{-0.5cm}
\end{figure}

Below 0.99 GeV the contribution from $2\pi$, using the CMD-3 data, is\,:
\beq
a_\mu^{2\pi}\vert^{hvp}_{lo}(\leq 0.99) =(5109\pm 35)\times 10^{-11}.
\label{eq:2pi-0.99}
\eeq
For an appropriate comparison with the ones quoted in Table\,\ref{tab:compare}, we add the contribution of the data from 0.993 to 1.2 GeV given by CMD-3.  The polynomials-fit of the data is given in Fig.\,\ref{fig:cmd3-1.2}.
This region gives:
\beq
a_\mu^{2\pi}\vert^{hvp}_{lo}(0.99-1.2) =(90.1\pm 1.5)\times 10^{-11},
\label{eq:2pi-0.99-1.2}
\eeq
\subsection*{ \hspace*{0.5cm} \d  Comparison of different  results}
A comparison of the  estimates from different experiments ($e^+e^-$ and $\tau$-decays) are shown in Table\,\ref{tab:compare} (see also Ref.\,\cite{MASJUAN}). . 

   {\footnotesize
\begin{table}[hbt]
\setlength{\tabcolsep}{1.pc}
  \begin{center}
    { \footnotesize
  \begin{tabular}{| llll |}
\hline
\rowcolor{yellow}Experiment&\boldmath$\sqrt{t}$\,\bf [GeV] &\boldmath$a^{2\pi}_\mu\vert^{hvp}_{l.o}\times 10^{11} $&Refs.\\ 
 \hline 
\boldmath$e^+e^-$ & 1.2 &\boldmath$5199\pm 35$ &\bf CMD-3 This work \\
&1.97&$5035\pm 19$&\cite{NOMURA} \\
&$M_\tau$&$5079\pm 34$&\cite{DAVIER2}\\
\boldmath$\tau$\bf -decays &$M_\tau$&$5110\pm 56$&ALEPH\,\cite{DAVIER}\\
&$M_\tau$&$5269\pm122$ &OPAL\,\cite{OPAL,DAVIER} \\
&$M_\tau$&$5131\pm 58$&CLEO\,\cite{CLEO}\\
&$M_\tau$& $5235\pm 39$&BELLE\,\cite{BELLEb}\\
&$M_\tau$ &\boldmath$5183\pm 27$&\bf Mean \boldmath $\tau$\bf-decays\\
   \hline
  
 
\end{tabular}}
 \caption{Comparison of the estimates of
  $a_\mu^{2\pi}\vert^{hvp}_{l.o}$ from different sources and different cuts of $\sqrt{t}$.} \label{tab:compare} 
 \end{center}
\end{table}
} 

-- One can notice from the Table that, despite the low cut $\sqrt{t}=1.2$ GeV used, the  estimate from the CMD-3 
data is larger than the two most precise ones\,\cite{NOMURA,DAVIER2}
from $e^+e^-$ data. 
 
 -- One may also remark that the uncertainties given by Ref.\,\cite{NOMURA} are much smaller than the other determinations. 
 
 -- The results from $\tau$-decay taken up to the $\tau$-mass are in a better agreement with our result taken from the data below $\sqrt{t}$ 1.2 GeV but these results should be lower if one adds, to our result, the contribution from 1.2 GeV to $M_\tau$. 
\subsection*{ \hspace*{0.5cm} \b Comparison with some Lattice calculations}
\begin{figure}[H]
\begin{center}
\hspace*{0.1cm}\includegraphics[width=12cm]{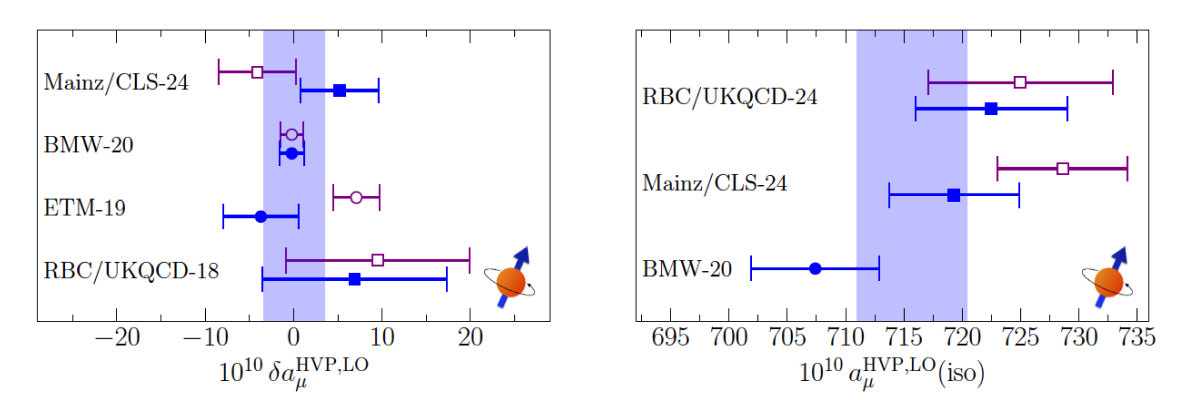}
\vspace*{-0.5cm}
\caption{\footnotesize Different lattice calculations reviewed in Ref.\,\cite{WP25}: Left: Isospin breaking;  Right: SU(2) rotation.}
\label{fig:lattice}
\end{center}
\vspace*{-0.5cm}
\end{figure}

There are some progresses in the lattice calculations but the errors and the range spanned by the results from different groups are still too large for the results to be conclusive (see Fig.\ref{fig:lattice}). 

Instead of averaging these results like in\,\cite{WP25}, we consider for a comparison the most precise one from  the BMW group\,\cite{BMW}:
\beq
a_\mu\vert^{hvp}_{lo}\vert_{bmw}= 7072 (55)\times 10^{-11},
\label{eq:bmw-latt}
\eeq
where one can see a good agreement with our determination in Eq.\,\ref{eq:amu-res}.

\section{Some other hadronic contributions from phenomenology}
\subsection*{ \hspace*{0.5cm} \b Hadronic vacuum polarization  to higher orders}
The NLO and N2LO contributions are known and reviewed in Ref.\,\cite{CALMET,WP25}. The estimate from $e^+e^-\to $ Hadrons data is:
\beq
\Delta a_\mu\vert^{hvp}_{nlo}= -(99.6\pm 1.3)\times 10^{-11},~~~~~~~~~~~~~~\Delta a_\mu\vert^{hvp}_{n2lo}= (12.4\pm 1.0)\times 10^{-11},
\label{eq:hvp-ho}
\eeq
\subsection*{ \hspace*{0.5cm} \b Light by light scattering}
Estimates of different meson contributions are reviewed in Ref.\,\cite{WP25} where the $I=0$ scalar meson contribution has been estimated in Ref.\,\cite{KNECHT}. They read:
\beq
a_\mu\vert^{hlbl}_{lo}= (103.3\pm 8.8)\times 10^{-11},~~~~~~~~~~~~~~\Delta a_\mu\vert^{hlbl}_{nlo}= (2.6\pm 0.6)\times 10^{-11},
\label{eq:hlbl-pheno}
\eeq
For a comparison, we give the Lattice averaged estimate\,\cite{WP25}:
\beq
a_\mu\vert^{hlbl}_{lo}\vert_{lattice}= (122.5\pm 9.0)\times 10^{-11}.
\label{eq:hlbl-lattice}
\eeq
\section{Total sum of  the hadronic contributions and SM predictions of $a_\mu$}
\subsection*{\hspace*{0.5cm}\b Phenomenology}
Adding the previous hadronic contributions from phenomenological approaches: 

-- Hadronic vacuum polarization  @ LO in Eq.\,\ref{eq:amu-res} and  @ NLO $\oplus$ N2LO in Eq.\,\ref{eq:hvp-ho},

-- Hadronic Light by Light scattering @ LO and NLO) , 

we obain  the sum of the hadronic contributions estimated
phenomenologically in Eq.\,\ref{eq:hlbl-pheno}\,:
\beq
a_\mu^h\vert_{pheno} = (7112\pm 38)\times 10^{-11}. 
\eeq
Comparing this value with the one in Eq.\,\ref{eq:amuth}, one deduces:
\bea
\Delta a_\mu^{pheno}\equiv \la a_\mu^{exp}\ra- a_\mu^{pheno} &=& (85\pm 14.5_{exp} \pm 4.5_{qed\oplus ew}\pm 38.2_{hadrons})\times 10^{-11}, \nnb\\
&=& (85\pm 41)\times 10^{-11}. 
\label{eq:amu-pheno}
\eea
\subsection*{\hspace*{0.5cm}\b Lattice}
It is informative to compare the previous phenomenological results with the one from the most precise lattice calculation of the LO hadronic vacuum polarization in Eq.\,\ref{eq:bmw-lattice} and the available result of the hadronic light by light scattering in Eq.\,\ref{eq:hlbl-lattice} to which we add the higher order contributions to the hadronic vacuum polarization in Eq.\,\ref{eq:hvp-ho}. We deduce:
\beq
a_\mu^h\vert_{lattice} = (7072\pm 55)\times 10^{-11}. 
\eeq
Using this value  in Eq.\,\ref{eq:amuth}, one deduces:
\bea
\Delta a_\mu^{lattice}\equiv \la a_\mu^{exp}\ra- a_\mu^{lattice} &=& 90\pm 14.5_{exp} \pm 4.5_{qed\oplus ew}\pm 55_{lattice} \nnb\\
&=& (90\pm 56)\times 10^{-11}. 
\label{eq:amu-lattice}
\eea
This result differs from Ref.\,\cite{WP25} as we have not attempted to average the lattice results for HVP at LO where  most of them are inaccurate. We have not also averaged  the HLbL results from phenomenology and lattice. 

\subsection*{\hspace*{0.5cm}\b Attempt to give an unique SM prediction}
As the two results in Eqs.\,\ref{eq:amu-pheno} and \ref{eq:amu-lattice} are consistent within the errors, we attempt to take their average:
\beq
\Delta a_\mu^{SM}\equiv \la a_\mu^{exp}\ra- a_\mu^{SM} = (87\pm 33)\times 10^{-11},
\eeq
which we consider as a final estimate of $a_\mu^{SM}$.  Then, we conclude that the estimate of the Hadronic vacuum polarization using $e^+e^-\to$ Hadrons data using the new CMD-3 data indicates a deviation of the SM prediction from the FNAL data by about $2.1\,\sigma$,  while the most precise Lattice result indicates a smaller deviation of about $1.6\,\sigma$. A tentative average of the two results leads to a $2.6\,\sigma$ deviation. 

\section{The LO Hadronic Vacuum Polarization contribution to the $\tau$  anomaly  $a_\tau\vert^{hvp}_{lo}$}
We extend the previous analysis to the $\tau$ anomaly. Due to the larger value of $M_\tau$ compared with $m_\mu$, the relative weight
of the low energy region in the integral (Eq.\,\ref{eq:amu-form}) decreases.  We have not redone the analysis in Ref.\,\cite{SNe} but only compile them  in Table\,\ref{tab:amu1} as the effect of the new fit of the data is almost negligible for the case of the muon.  Adding the estimate of the $P\gamma$ process of $(21\pm 0.4)\times 10^{-9}$ from\,\cite{NOMURA}, we deduce
\beq
a_\tau\vert^{hvp}_{lo}=(3516\pm 25)\times 10^{-9}.
\eeq
 It is remarkable to compare this result with the 1st estimate\,\cite{SN78}:
\beq
a_\tau\vert^{hvp}_{lo}=(3700\pm 400)\times 10^{-9},
\eeq
where the accuracy has increased by a factor 16 thanks to the experimental effort for measuring the cross-section of the $e^+e^-\to$ Hadrons process. 
\section{The LO Hadronic Vacuum Polarization contribution to $\Delta\alpha^{(5)}(M_Z^2)$}
This contribution is defined as:
\beq
\Delta \alpha_{had}^{(5)} (M_Z^2)=-\ga \frac{\alpha}{3\pi}\dr M^2_Z\int_{4m_\pi^2}^\infty \frac{R_{ee}(t)}{t(t-M_Z^2)},
\eeq
where $R_{ee}$ is the ratio of the $e^+e^-\to$ Hadrons over the  $e^+e^-\to \mu^+\mu^-$ total cross-sections and the upper index (5) corresponds to 5 flavours.
The contributions from different regions are shown in Table\,\ref{tab:amu1} where at the  $Z_0$ pole, we take the principal value of the integral\,:
\beq
\int_{4m_\pi^2}^{(M_Z- \Gamma_Z/2)^2}\hspace*{-1cm} dt\,f(t)+\int_{(M_Z+ \Gamma_Z/2)^2}^{\infty}\hspace*{-1cm}dt\,f(t),
\eeq
where $\Gamma =2.5$ GeV is the total hadronic $Z$-width. We add to the QCD continuum contribution the one of the $Z$-pole estimated to be\,\cite{YND17}:
\beq
\Delta \alpha_{had}^{(5)}\vert_{M_Z}= 29.2\times 10^{-5}.
\eeq
Then, we obtain the total sum given in Table\,\ref{tab:amu1}, to which, we add the $P\gamma$ contribution of $(4.4\pm 0.1)\times 10^{-5}$:
\beq
\Delta \alpha_{had}^{(5)} (M_Z^2)= (2770.7\pm 4.5)\times 10^{-5},
\eeq 
which improves and confirms our previous determination in Ref.\,\cite{SNalfa} (for a review, see e.g. Ref.\,\cite{JEGER}). 
\section{Conclusions}
\b We have revisited our previous determination of the lepton anomalies and $\Delta\alpha^{(5)}_{hadrons}(M_Z^2)$ in Ref.\cite{SNe} by re-examining the different fitting procedure of the CMD-3 data\,\cite{CMD3}
and by replacing the NWA used for the $\omega$ by the $e^+e^-\to 3\pi$ most precise data of BABAR\,\cite{BABAR} and recent one of BELLE II\,\cite{BELLE}.  We notice that our results in Ref.\,\cite{SNe} are almost unaffected by these new fits, but one has to add the contribution from $\pi^0\gamma$ and $\eta\gamma$ which are not included in the CMD-3 data\,\cite{IGNATOV} not taken into account in Ref.\,\cite{SNe}. 

\b We found that within the recent precise data of $a_\mu$ from FNAL the estimate from $e^+e^-\to$ hadrons of the LO Hadronic Vacuum Polarization (HVP) indicates a deviation from the Standard Model (SM) prediction by $2.1\sigma$ compared to the most precise lattice one of $1.6\sigma$. 

\b To have a sharper conclusion, the $e^+e^-\to Hadrons$  data still need to be improved in the  region below the $\rho$-meson mass which is the main source of errors (see Table\,\ref{tab:amu1}), while the lattice calculations, in order  to be more significant should have a much better accuracy than presently available.  

\b New experimental projects such as the MUonE experiment  aims to measure directly the LO-HVP contribution to $a_\mu$ while the E-34 experiment plans to provide a new independent measurement of $a_\mu$.

\section{Acknowledgements}
It is a pleasure to thank A. Driutti, A. Pich and E. de Rafael  for stimulating questions and discussions. 


\end{document}